\documentclass[reprint,aps,pre,showpacs]{revtex4-2}
 \usepackage{graphicx}
 \usepackage{amsfonts}
 \usepackage{natbib}
 \usepackage{amsmath}
\usepackage{amssymb}
\usepackage{enumerate}
\usepackage[colorlinks,citecolor=blue,linkcolor=blue]{hyperref} 
 \usepackage{mathtools}
 \begin{document}
  \title{Chaos and quantization of the three-particle generic
  Fermi-Pasta-Ulam-Tsingou model II: phenomenology of quantum eigenstates}

  \author{Hua Yan}
  \email{yanhua@ustc.edu.cn}
  \affiliation{CAMTP - Center for Applied Mathematics and Theoretical
  Physics, University of Maribor, Mladinska 3, SI-2000 Maribor, Slovenia,
  European Union}
  \author{Marko Robnik}
  \email{Robnik@uni-mb.si}
  \affiliation{CAMTP - Center for Applied Mathematics and Theoretical
  Physics, University of Maribor, Mladinska 3, SI-2000 Maribor, Slovenia,
  European Union}

  \date{\today}

\begin{abstract}
    We undertake a thorough investigation into the phenomenology of quantum eigenstates,  in the three-particle FPUT model. Employing different Husimi functions, our study focuses on both the $\alpha$-type, which is canonically equivalent to the celebrated H\'enon-Heiles Hamiltonian, a nonintegrable and mixed-type system, and the general case at the saddle energy where the system is fully chaotic. Based on Husimi quantum surface of sections (QSOS), we find that in the mixed-type system, the fraction of mixed eigenstates in an energy shell $[E-\delta E/2, E+\delta E/2]$ with $\delta E\ll E$ shows a power-law decay with respect to the decreasing Planck constant $\hbar$. Defining the localization measures in terms of the R\'enyi-Wehrl entropy, in both the mixed-type and fully chaotic systems, we find a better fit with the beta distribution and a lesser degree of localization,  in the distribution of localization measures of chaotic eigenstates, as the controlling ratio $\alpha_\mathcal{L} = t_H /t_T$ between the Heisenberg time $t_H$ and the classical transport time $t_T$ increases. This transition with respect to $\alpha_\mathcal{L}$ and the power-law decay of the mixed states, together provide supporting evidence for the principle of uniform semiclassical condensation (PUSC) in the semiclassical limit. Moreover, we find that in the general case which is fully chaotic, the maximally localized state, is influenced by the stable and unstable manifold of the saddles (hyperbolic fixed points), while the maximally extended state notably avoids these points, extending across the remaining space, complementing each other.
\end{abstract}
\pacs{01.55.+b, 02.50.Cw, 02.60.Cb, 05.45.Pq, 05.45.Mt}
\maketitle

\section{Introduction}
\label{sec1}
This paper is a continuation of our recent paper (paper I) \cite{yan2024chaosPaperI} in which we have studied in detail classical and quantum aspects of the well known paradigmatic Fermi-Pasta-Ulam-Tsingou (FPUT) model \cite{fermi1955studies}. We are specializing on the few-body case, the three-particle FPUT system, which is reduced to one particle in two-dimensional potential, described as two identical harmonic oscillators coupled by cubic and quartic potential terms. This is so called $\alpha$-FPUT (system) in the case of purely cubic couplings (it is precisely the celebrated H\'enon-Heiles system \cite {henon1964applicability}), or $\beta$-FPUT (in the case of purely quartic couplings), and $\alpha\beta$-FPUT
in the general case. 

Our focus in paper I was the relevance of classical chaos, in particular of the mixed-type structure of the phase space and the degree of chaos as measured by the smaller alignment index (SALI) method \cite{skokos2001alignment,skokos2003does,skokos2004detecting,bountis2012complex,skokos2016chaos}, for the quantal energy spectra and their statistical properties. We analyzed the semiclassical limiting behavior of the energy spectra. First, our explicit analytic results for the quantum density of states (DOS), using the Thomas-Fermi rule, are derived and compared with the exact (numerical) quantum DOS. They perfectly agree, as well as they do agree with the DOS based on the quantum typicality method \cite{de2004computational,weisse2006kernel,bartsch2009dynamical}. Next, the energy spectra were studied, in the case of mixed-type classical (divided) phase space. In the case of no quantum (or dynamical) localization the Berry-Robnik (BR) picture \cite{berry1984semiclassical} is confirmed, the level spacing distribution follows the BR formula: The extracted quantum BR parameter (the relative size of the chaotic component) agrees with the classical value within better than one percent. If the localization effects of the chaotic eigenstates are included, the level spacing distribution is generalized to Berry-Robnik-Brody (BRB) distribution, as the localized chaotic eigenstates are (empirically) well captured by the Brody level spacing distribution \cite{prosen1993energy,prosen1994semiclassical,batistic2010semiempirical}.
The BRB distribution is found to excellently fit the numerical distribution. In this way we have followed the classical and quantum transition from regularity to chaos, by increasing the energy of the system. This transition
also has been described by the ratio of consecutive level spacings (so-called spacing ratio, which does not depend on the unfolding procedure). 

In the present paper we study in detail the structure and statistical properties of the eigenstates, by means of the Husimi functions, which are Gaussian smoothed Wigner functions, or equivalently, absolute squared projections of the eigenstates on the coherent states. We study the quantum Husimi surfaces of section (QSOS) \cite{bogomolny1992distribution,prosen1996quantum}, which are precise analog of the classical Poincar\'e SOS, as well as various projections of the Husimi functions. In doing so, the classical structure of the mixed-type phase space is clearly revealed. We use the overlap index $M$, introduced by Batisti\'c and Robnik \cite{batistic2013dynamical}, which measures the degree of the overlap of the QSOS with the clasical regular $(M=-1)$ or chaotic $(M=+1)$ region, or it measures the partial overlap with both regions such that $|M|<1$. Using this index we look at its distribution at given energy (in a very narrow energy shell, for the first time, in a genuinely continuous Hamiltonian system with a smooth
potential) and sufficiently small Planck constant, comprising about 1200 eigenstates. We show again that the fraction of mixed states with $|M|<1$ decreases as a power law with decreasing Planck constant, like it was shown before in billiards \cite{lozej2022phenomenology}, kicked top \cite{wang2023power} and Dicke model \cite{wang2023mixed}. 

Using the index $M$ we separate the chaotic eigenstates $M\ge0.8$ (those which live inside the classical chaotic component,
and can be localized there) and study their quantum phase space localization phenomena. We introduce various entropy
localization measures (ELM) based on $\alpha$-R\'enyi-Wehrl entropy: The information entropy at $\alpha=1$ and inverse participation ratio at $\alpha=2$. Then we study the statistical properties of ELM, showing that their distribution is well captured by the beta distribution, in perfect agreement with the results of the same analysis in many other systems, e.g. billiard systems  \cite{batistic2019statistical,lozej2021effects,lozej2022phenomenology,lozej2022quantum}, kicked top \cite{wang2023statistics} and Dicke model \cite{wang2020statistical}. We also observe that most localized eigenstates in the fully chaotic (ergodic) general $\alpha\beta$-FPUT system, with  $\lambda=1/16$ at the energy $E=1/3$, are strongly localized around the hyperbolic fixed point and its stable and unstable manifolds, contributing to the lower tail of the localization measure distribution. This is an analogy of the strong localization in the classical stickiness regions observed in the ergodic lemon billiard \cite{lozej2021effects}. This effect, however, is expected to disappear in the strict semiclassical limit, at smaller Planck constant, according to PUSC. 

The paper is structured as follows. In Sec. \ref{sec2} we define the details of the classical FPUT system, specialize to the three particle case and perform the quantization procedure by introducing the rotated bosonic operators and the corresponding circular two-mode basis and relating them to the Cartesian-mode basis. In Sec. \ref{sec3} we define the Husimi functions, the quantum surface of section, i.e. Husimi QSOS,  as well as the various projections of the Husimi function. In Sec. \ref{sec4} we introduce the overlap index $M$, look at its distributions and demonstrate the power-law decay of the fraction of the mixed eigenstates with the decreasing Planck constant. In Sec. \ref{sec5} we introduce the localization measures based on the R\'enyi-Wehrl entropy, and look at their distribution, also in relation to the controlling parameter, the ratio of the Heisenberg time and the classical transport time. In Sec. \ref{sec6} we conclude and discuss the results and the outlook. In the Appendices \ref{appA}-\ref{appD} we present mathematical details, as well as the gallery of eigenstates, represented by the Husimi functions both for the H\'enon-Heiles system and the general FPUT system.

\section{Hamiltonian and circular two-mode basis}
\label{sec2}
In this section, we present a brief overview of the canonically equivalent Hamiltonian that governs the classical dynamics of a three-particle FPUT, and its quantization using rotated bosonic operators. Then, for the discussion of properties of circular two-mode basis, in which the matrix form of Hamiltonian is expressed, we derive a Wigner $d$-matrix decomposition of the unitary transformation from this circular basis to the Cartesian two-mode basis. 

\subsection{Canonically Equivalent Hamiltonian of generic three-particle FPUT}
\label{sec2.1}
\begin{figure}
    \includegraphics[width=1\linewidth]{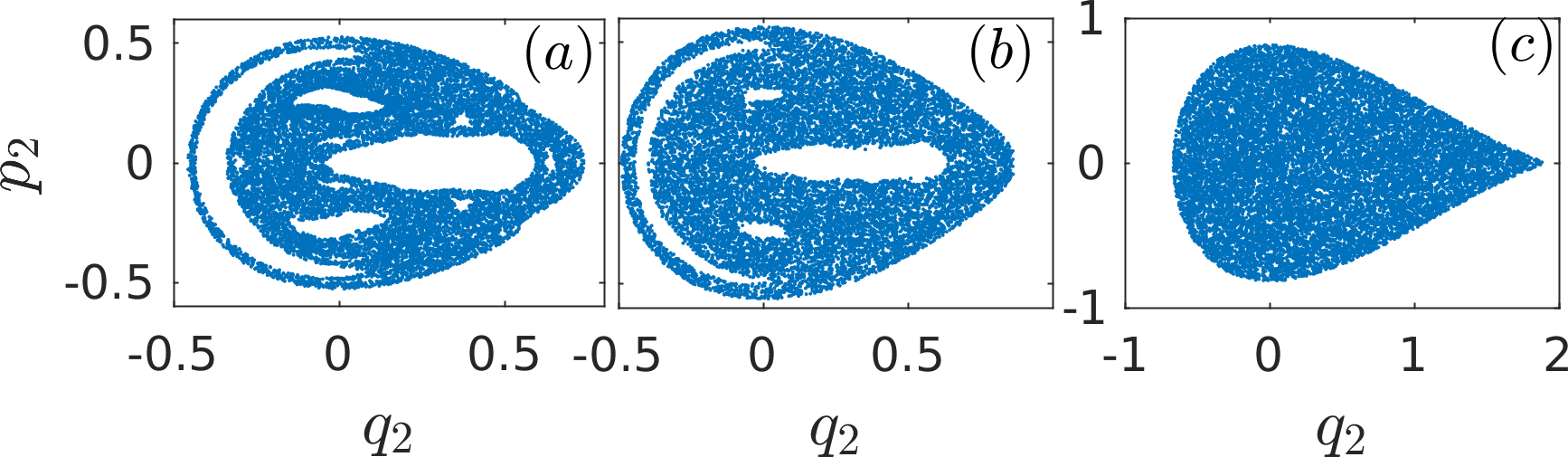}
    \caption{Classical Poincar\'e surface of section for $\alpha$-FPUT with $\alpha=1$ at two energies, $E=0.14$ ($a$) and $E=0.16$ ($b$), and for the general case with $\lambda=1/16$ at $E=1/3$ the saddle energy $(c)$, generated by one single chaotic orbit.}
    \label{fig:HenonPSS}
\end{figure}

The generic three-particle FPUT is a chain of $N=3$ moving particles with nearest neighbor interaction given by a potential $V$,  the Hamiltonian of such a system is given by
  \begin{align}
    \label{eq:fpu-pbc}
    H=\sum_{j=1}^N\left(\frac{y_i^2}{2} + V(x_{j+1}-x_j)\right),
  \end{align}
with the periodic boundary condition (PBC) $x_1=x_{N+1}$, where $x_j$ is the displacement of the particles with respect to the equilibrium positions and $y_j$ is the corresponding momentum, and the potential is 

  \begin{align}
    V(s)=\frac{1}{2}s^2+\frac{\alpha}{3}s^3+\frac{\beta}{4}s^4.
  \end{align}
 $\alpha$-FPUT  refers to the case $\beta=0$, while in the  $\beta$-FPUT case  $\alpha=0$. In paper I, after transforming to the normal mode representation, we have rescaled the Hamiltonians in the center of mass frame and obtained the canonically equivalent Hamiltonians  
  \begin{align}
    \label{eq:helon-heiles}
    H_\alpha = \frac{1}{2}\sum_{i=1}^2(p_i^2+q_i^2)+\alpha(q_1^2q_2-\frac{1}{3}q_2^3)
\end{align}
for $\alpha$-FPUT, which is also known as the H\'enon-Heiles Hamiltonian, and
\begin{align}
    \label{eq:beta-FPUT}
    H_\beta=\frac{1}{2}\sum_{i=1}^2(p_i^2+q_i^2)+\frac{3\beta}{4}(q_1^2+q_2^2)^2
\end{align}
for $\beta$-FPUT, while for the general case ($\alpha\beta$-FPUT)
\begin{align}
    \label{eq:alpha-beta-FPUT}
    H = \frac{1}{2}\sum_{i=1}^2(p_i^2+q_i^2)+ q_1^2q_2-\frac{1}{3}q_2^3+ \lambda(q_1^2+q_2^2)^2,
\end{align}
where we denote the coupling parameter of the quartic term as $\lambda=3\beta/(4\alpha^2)$. In Fig. \ref{fig:HenonPSS} we show several examples of classical Poincar\'e surface of section (SOS). 

For the quantization, we define $\hat{q}_\pm = \hat{q}_1\pm i\hat{q}_2, \
\hat{p}_{\pm}=\hat{p}_1\pm i\hat{p}_2$, and introduce the rotated bosonic operators as
\begin{equation}
\label{eq:rotated-bosonic}
\begin{aligned}
   a_\pm &= \frac{1}{\sqrt{2}}(a_1\mp i a_2)=\frac{1}{2\sqrt{\hbar}}(\hat{q}_\mp +i\hat{p}_\mp) ,\\
   a_\pm^\dagger &= \frac{1}{\sqrt{2}}(a_1^\dagger \pm i a_2^\dagger)= \frac{1}{2\sqrt{\hbar}}(\hat{q}_\pm - i\hat{p}_\pm), 
\end{aligned}
\end{equation}
where the annihilation and creation operator fulfill the canonical commutation relations $[a_i,a_j^\dagger]=\delta_{ij}$, with $i,j\in\{1,2,\pm\}$. We can then check that for $\hat{n}$ the number operator and the angular momentum operator $\hat{\ell}=(\hat{q}_1\hat{p}_2-\hat{q}_2\hat{p}_1)/\hbar$ fulfill
\begin{equation}
\begin{aligned}
    &\hat{n}=a_1^\dagger a_1+a_2^\dagger a_2=a_+^\dagger a_++ a_-^\dagger a_-,\\
    &\hat{\ell}= i(a_2^\dagger a_1 -a_1^\dagger a_2)=a_+^\dagger a_+-a_-^\dagger a_-,
\end{aligned} 
\end{equation}
and the quantized Hamiltonian of $\alpha$-FPUT can be compactly written as
\begin{align}
    \hat{H}_\alpha=\hbar(\hat{n}+1)-i\alpha(\hat{q}_+^3-\hat{q}_-^3)/6,
\end{align}
while for the general case
\begin{align}
    \hat{H}=\hbar(\hat{n}+1)-i(\hat{q}_+^3-\hat{q}_-^3)/6+\lambda(\hat{q}_+\hat{q}_-)^2.
\end{align}

In circular two-mode basis states $|n,l\rangle$, defined as the simultaneous eigenfunctions of  $\hat{n}$ and $\hat{\ell}$, where $l=-n,-n+2,\cdots,n\  (n\in \mathbb{N}_0)$, $n$ is the radial quantum number and $l$ is the orbital angular momentum (OAM), we have the expressions for matrix elements of the cubic and quartic coupling terms (see Appendix \ref{appA}). The cubic term in quantum $\alpha$-FPUT couples states with $\Delta l=\pm 3$, therefore the coupling takes place in three decoupled sets of basis states: the singlet $\{a\}$ $l\in \{\dots -6, -3 ,0, 3, 6\dots\}$, two doublets with the same eigenspectra $\{b\}$ $l\in \{\dots -5, -2 ,1, 4\dots\}$ and $\{c\}$ $l\in \{\dots -4, -1 ,2, 5\dots\}$. This property of eigenspectra can also be verified from the $C_{3v}$ symmetry of the classical Hamiltonian, that they must belong to the irreducible representations of the point symmetry group $C_{3v}$: the subspaces of two doublets are of $E$ symmetry and degenerate, and the singlet is a combination of ($A_1$, $A_2$) symmetry. The quartic term in quantum $\beta$-FPUT does not introduce any coupling between different $l$, due to the conservation of angular momentum.

\subsection{Properties of the circular two-mode basis}
\label{sec2.2}
Given the definition of rotated-bosonic operators and the corresponding circular two-mode basis $|n,l\rangle$, the Fock space can be spanned as
\begin{align}
    \label{eq:fock-basis}
    |n,l\rangle =|n_++n_-,n_+-n_-\rangle = \frac{(a_+^{\dagger})^{n_+} (a_-^\dagger)^{n_-}}{\sqrt{n_+!n_-!}}|00\rangle,
\end{align}
where $|00\rangle$ is two-mode vacuum and $n=n_++n_-,\  l=n_+-n_-$. The interpretation of these relations is direct: 
\begin{align}
\hat{n}_+=a_+^\dagger a_+,\  \hat{n}_-=a_-^\dagger a_-,\ \hat{\ell}=\hat{n}_+-\hat{n}_-,
\end{align}
$\hat{n}_+$ stands for counterclockwise rotation with positive angular momentum, and $\hat{n}_-$ refers to clockwise rotation with negative angular momentum.

The expansion of Eq. \eqref{eq:fock-basis} defines the unitary transformation $\Omega$ between the circular-mode basis $|n,l\rangle $ and the Cartesian-mode basis $|n_1,n_2\rangle$ 
\begin{align}
    \label{eq:unitary-m}
   |n,l\rangle
   = \sum_{n_1,n_2}\Omega_{n_1n_2}^{nl}|n_1,n_2\rangle,\ n=n_1+n_2,
\end{align}
where $\Omega^{nl}_{n_1n_2}$, the elements of  $\Omega$,  are all sums of many factorials of large numbers with alternating signs for $n\gg 1$, because of which a direct numerical evaluation would suffer from serious errors and instability at large $n$, very similar to the longstanding problem of the evaluation of the Wigner $d$-matrix at high spins \cite{tajima2015analytical}. 

To avoid these large factorials in the numerical evaluation, a further analytical expression of $\Omega^{nl}_{n_1n_2}$ is essential.  Setting $n=2j, l=2m$, in Appendix \ref{appB} we have proven that 
\begin{align}
    \label{eq:unitary-wigner}
    |n,l\rangle :=|2j,2m\rangle =\sum_{m'=-j}^{j}O_{m,m'}^j|j+m',j-m'\rangle,
\end{align}
with the correspondence 
\begin{align}
    \label{eq:co-coeff}
    \Omega_{n_1n_2}^{nl}=O_{m,m'}^j, \ n_1=j+m',\  n_2=j-m',
\end{align}
where the coefficients $O_{m,m'}^j$ are particular instances of the rotation matrix elements, which we write as
\begin{align}
    O_{m,m'}^j &= i^{m'-j}d_{m,m'}^j(\pi/2)\nonumber\\
    &=i^{m'-j}\langle jm|e^{-i\pi J_y/2}|jm'\rangle,
\end{align}
where $d_{m,m'}^j(\theta)$ is the Wigner $d$-matrix, $J_y$ is one generator of the Lie algebra of SU(2) and SO(3), component of the angular momentum operator. $O_{m,m'}^j $ then can be calculated quite effectively by the sparse matrix exponential, using the kernel polynomial method \cite{weisse2006kernel}.

\section{Husimi function}
\label{sec3}
As a Gaussian coarse graining of the Wigner function, Husimi function is a powerful tool to study the quantum-classical correspondence in quantum systems. Equivalently, it can be defined as the projection of the wave function onto the coherent state. In quantum three-particle FPUT, for the $k$-th energy eigenstate $|E_k\rangle$, the Husimi function is given by 
\begin{align}
    \mathcal{H}_k(\alpha_1,\alpha_2)\equiv \mathcal{H}_k(q_1,p_1;q_2,p_2)=|\langle \alpha_1,\alpha_2|E_k\rangle|^2,
\end{align}
where $|\alpha_1,\alpha_2\rangle = |\alpha_1\rangle \otimes |\alpha_2\rangle$ is the product coherent state, and
\begin{align}
    \frac{1}{\pi^2}\int|\alpha_1,\alpha_2\rangle\langle \alpha_1,\alpha_2|d^2\alpha_1d^2\alpha_2= I.
\end{align}
The coherent state $|\alpha_i\rangle$ defined as $a_i|\alpha_i\rangle=\alpha_i|\alpha_i\rangle $ with $\alpha_i=\sqrt{1/2\hbar}(q_i+ip_i)$, can be expanded in terms of Fock states as follows
\begin{align}
    |\alpha_i\rangle=e^{-\frac{1}{2}|\alpha_i|^2}\sum_{n_i=0}^\infty\frac{\alpha_i^{n_i}}{\sqrt{n_i!}}|n_i\rangle, \ i=1,2.
\end{align}
 The Husimi function for  $|E_k\rangle$ can be then written as
 \begin{equation}
    \begin{aligned}
        \label{eq:husimi}
        \mathcal{H}_k(\alpha_1,\alpha_2)=|\sum_{n_1,n_2} B_{n_1n_2}^k\langle\alpha_1|n_1\rangle\langle\alpha_2|n_2\rangle|^2 ,\\
        B_{n_1n_2}^k=\sum_{n,l} C_{nl}^k\Omega_{n_1n_2}^{nl}=\sum_{j,m,m'}C_{jm}^k O_{m,m'}^j,
    \end{aligned} 
 \end{equation}
where $C_{jm}^k:=C_{nl}^k =\langle n,l|E_k\rangle $ are the expanding coefficients. From the properties of circular two-mode basis, we have verified  in Appendix \ref{appC}  that the Husimi function can be equivalently expressed as 
\begin{align}
    \label{eq:husimi-cart}
    \mathcal{H}_k(\alpha_1,\alpha_2)=|\sum_{n,l}C_{nl}^k \langle \alpha_+,\alpha_-|n,l\rangle|^2,
\end{align}
where  $\alpha_\pm =(\alpha_1\mp i\alpha_2)/\sqrt{2}$, and
\begin{align}
    \langle \alpha_+,\alpha_-|n,l\rangle:=\frac{(\alpha_+^*)^{n_+}(\alpha_-^*)^{n_-}}{\sqrt{n_+!n_-!}}e^{-(|\alpha_+|^2+|\alpha_-|^2)/2},
\end{align}
provides coherent state $|\alpha_\pm\rangle$ defined to be the eigenstate of the (rotated) annihilation operator $a_\pm$.

\subsection{Husimi Quantum Poincar\'e surface of section}
\label{sec3.1}
To facilitate a comprehensive comparison with classical calculations, especially the classical SOS, the quantum Poincar\'e surface of section (QSOS) based on Husimi function as the quantum analog of the classical SOS is necessary to be defined. Similar to the classical SOS on $q_1=0$ plane with $p_1>0$, one can define the (normalized) Husimi QSOS namely as
\begin{align}
    \label{eq:QSOS-unitary}
    \bar{\mathcal{Q}}_k(\alpha_2)=\frac{1}{\mathcal{A}_k}\mathcal{H}_k(\bar{\alpha}_{1},\alpha_2),\ 
    \bar{\alpha}_1=(\bar{q}_1+i\bar{p}_1)/\sqrt{2\hbar},
\end{align}
where $\mathcal{A}_k=\frac{1}{\pi}\int_{\mathcal{S}} \mathcal{H}_k(\bar{\alpha}_{1},\alpha_2)d^2\alpha_2$ is a normalization constant and $\mathcal{S}$ denotes the surface of section of the $q_1=0$ plane at $E_k$. $\bar{q}_1=0$, $\bar{p}_1=p_1^+(q_1=0,q_2,p_2,E_k)$ is the positive classical momentum obtained from the classical Hamiltonian at the surface of section, 
\begin{align}
    p_1^+ = \sqrt{2E_k-2H(q_1=0,p_1=0,q_2,p_2)}.
\end{align}
The resulting Husimi QSOS of $|E_k\rangle$ then reads
\begin{align}
    \bar{\mathcal{Q}}_k(\alpha_2)=\frac{1}{\mathcal{A}_k}|\sum_{n_1,n_2}B_{n_1n_2}^k\langle \bar{\alpha}_1|n_1\rangle\langle\alpha_2|n_2\rangle|^2.
\end{align}
It can also be obtained directly from the circular basis according to the expression given in Eq. \eqref{eq:husimi-cart} 
\begin{align}
    \label{eq:QSOS-direct}
 \bar{\mathcal{Q}}_k(\alpha_2) =\frac{1}{\mathcal{A}_k}|\sum_{n,l}C_{nl}^k \langle \bar{\alpha}_+,\bar{\alpha}_-|n,l\rangle|^2,
\end{align}
where $\bar{\alpha}_\pm =(\bar{\alpha}_1\mp i\alpha_2)/\sqrt{2}$. The Husimi QSOS proves valuable for investigating  quantum systems with classical correspondence. It can serve to unveil the phenomenology of quantum eigenstates in mixed-type systems, such as the study of statistical properties of the localization measure of chaotic eigenstates, and the clarification of the effects of stickiness in quantum states in billiard systems \cite{lozej2021effects,batistic2019statistical,lozej2022quantum}.

\begin{figure*}
    \includegraphics[width=0.9\linewidth]{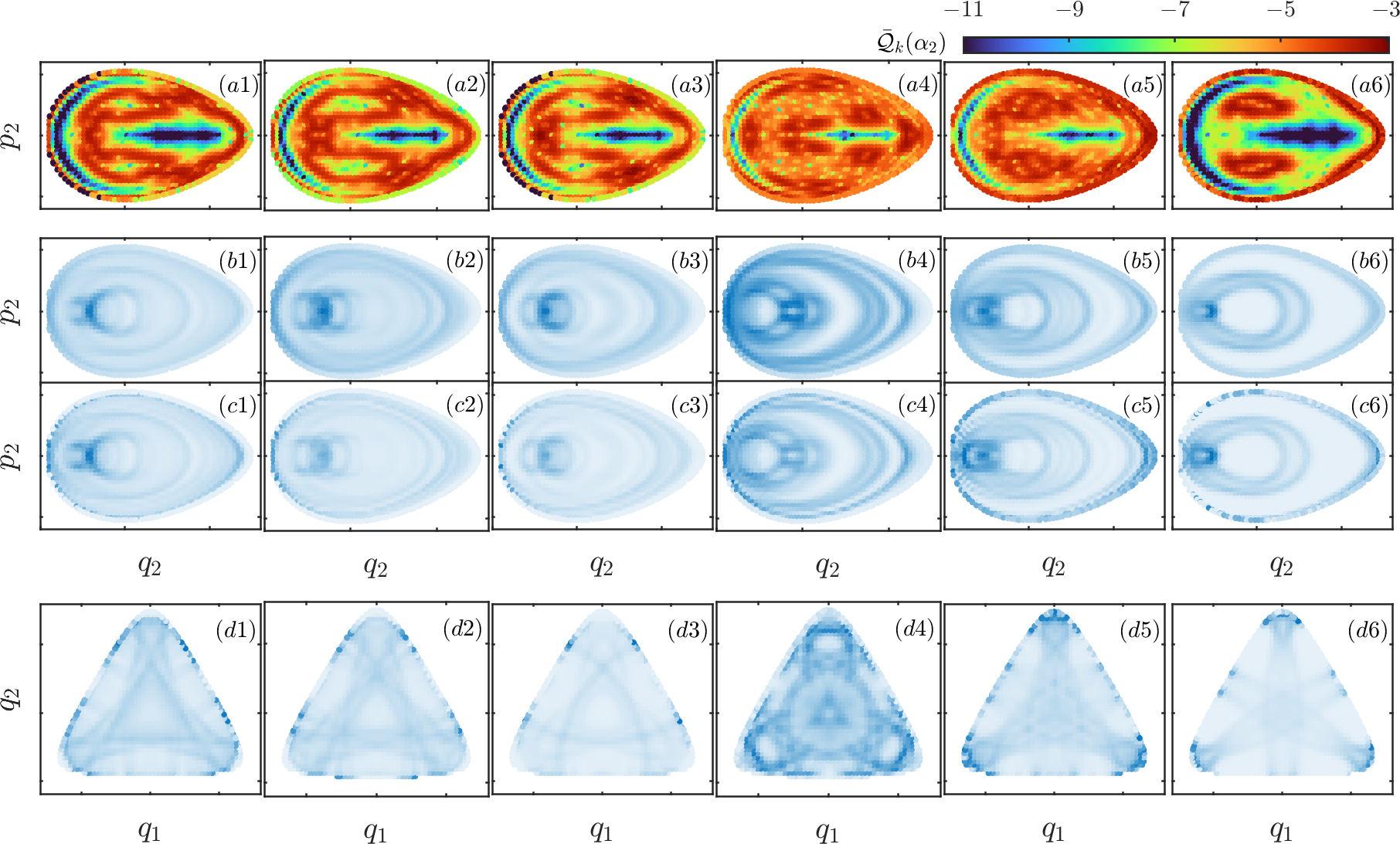}
    \caption{Examples of Husimi functions of six eigenstates in the energy interval $[E-\delta E/2,E+\delta E/2]$ at $E=0.14$ with $\delta E/E\simeq0.0014$, for the singlet of $\alpha$-FPUT with $\alpha=1$,  where the irreducible Hilbert space is of size $\mathcal{N}_S=60301$, with the cutoff of $n$ set to $N=600$ and the Planck constant $\hbar=1\times 10^{-3}$. Panels ($a1$)-($a6$): Husimi QSOS given by Eq. \eqref{eq:QSOS-direct} plotted in the logarithmic scale, the darkest blues show the area where $\bar{\mathcal{Q}}_k(\alpha_2)< 10^{-11}$. Panels ($b1$)-($b6$) are plots of the completely projected Husimi functions $\mathcal{P}_k(\alpha_2)$ of Eq. \eqref{eq:completeProj} and panels ($c1$)-($c6$) are the projected Husimi functions from the classical energy shell $\widetilde{\mathcal{P}}_k(\alpha_2)$ given by Eq. \eqref{eq:husimi-proj}. Panels $(d1)$-$(d6)$: projected Husimi functions in configuration space $\widetilde{\mathcal{P}}_k(q_1,q_2)$ of Eq. \eqref{eq:husimiProjConfig} plotted in linear scale. The darker colors in each panel for projected Husimi functions indicate  larger density, in a linear scale. Panels from the same column correspond to the same eigenstate.}
    \label{fig:HusimiEnsemble}
\end{figure*}

In Fig. \ref{fig:HusimiEnsemble}($a1$)-($a6$) we show the Husimi QSOS of six consecutive eigenstates in the energy interval $[E-\delta E/2,E+\delta E/2]$ at $E=0.14$ with $\delta E/E\simeq0.0014$, for the singlet of $\alpha$-FPUT with $\alpha=1$ (in all the following we set $\alpha=1$ as default), where $N$ the cutoff of $n$ is 600 and $\hbar=1\times 10^{-3}$.  The narrow energy window fulfills $\delta E/E\ll 1$, and this ensures that these consecutive levels are associated with the same classical SOS of $E=0.14$ in the semiclassical limit. From the semiclassical counting the number of energy levels, there are approximately 16 times more levels  provided above this energy for the numerical calculation, enough to achieve a good convergence of the numerical results with respect to the cutoff. A clear correspondence is observable when comparing plots of the Husimi QSOS with the classical SOS shown in Fig. \ref{fig:HenonPSS}($a$). For this mixed-type system, the displayed eigenstates do not solely concentrate on the invariant tori within the integrable region nor do they evenly distribute across the chaotic area: 
\begin{enumerate}[(i)]
    \item There are chaotic eigenstates with different degrees of localization, located almost totally in the chaotic region such as the ones shown in Fig. \ref{fig:HusimiEnsemble}($a1$)-($a3$), or flooding into the regular areas such as the cases shown in Fig. \ref{fig:HusimiEnsemble}($a4$)-($a5$).
    \item  Some eigenstates are strongly localized around regular regions as shown in Fig. \ref{fig:HusimiEnsemble}($a6$).   Surrounding the boundaries that separate the chaotic and regular region, visible patterns of Husimi QSOS manifest as a result of quantum tunneling. 
\end{enumerate}

One interpretation of these observations is that the Planck constant we have chosen here,  $\hbar=1\times 10^{-3}$, is not so deep in the semiclassical limit. In the present context, we have not shown eigenstates entirely confined within regular regions. A more comprehensive exploration of this phenomenon is reserved for detailed exposition in Section \ref{sec4}. Further numerous illustrative examples can be found in Appendix \ref{appD}, specifically showcased in Fig. \ref{fig:henonGallery}. These instances have been derived from a larger ensemble of eigenstates, a procedure undertaken after introducing the overlap index as a meaure to separate the chaotic and regular eigenstates. According to the principle of uniform semiclassical condensation (PUSC) of Wigner or Husimi functions \cite{shnirel1974ergodic,percival1973regular,berry1977regular,robnik1998topics,veble1999study,robnik2023recent}, the mixed states should disappear in the semiclassical limit according to a power law described below, leaving behind only regular and chaotic states.

\begin{figure}
    \includegraphics[width=0.9\linewidth]{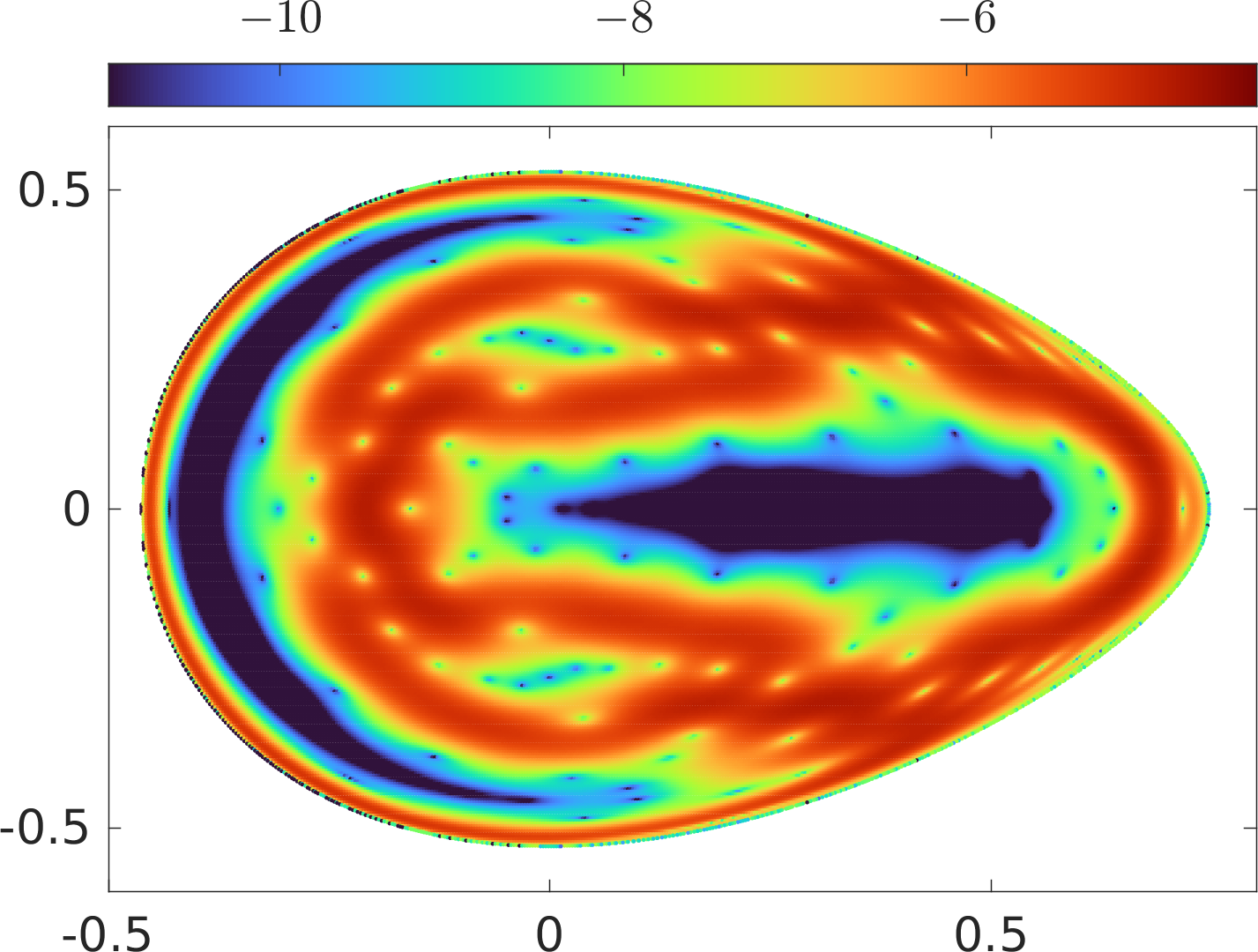}
    \caption{Husimi QSOS plotted in the logarithmic scale, of the same eigenstate as the one shown in Fig. \ref{fig:HusimiEnsemble}($a1$), but with much higher resolution about $\sim 1\times 10^{6}$ grid points. It corresponds to SOS of Fig. \ref{fig:HenonPSS}($a$).}
    \label{fig:detailQSOS}
\end{figure}

From Fig. \ref{fig:detailQSOS}, a replot with enhanced resolution of Fig. \ref{fig:HusimiEnsemble}($a1$), we have another important observation:  the non-uniformity (localization) of chaotic eigenstates is attributed to the classical boundaries between regular and chaotic regions, where variations in the magnitude of the Husimi QSOS are evident across these boundaries. Furthermore, within the chaotic region, there are embedded \emph{weak} (approximate) zeros of the Husimi QSOS \cite{leboeuf1990chaos,nonnenmacher1998chaotic,korsch1997zeros}, shown in  Fig. \ref{fig:detailQSOS} inside the light spots some magnitude smaller than the surroundings, which also contribute to the non-uniformity. These weak zeros are owing to the mathematical property of Husimi functions as the square moduli of complex analytic functions, also referred to as the Bargmann-Husimi representation of quantum states. Further factorization  of the Bargmann transform \cite{bargmann1961hilbert,bargmann1967hilbert} proves that the Husimi
functions are completely determined by its zeros, therefore the geometry of the quantum state can be essentially described by the distribution of the zeros,
the so-called stellar representation of quantum states. The distributions of zeros show contrasting behaviors for regular and chaotic states: they form regular ordered patterns along the lines of invariant curves for regular states and exhibit more disordered patterns for chaotic states. Qualitatively, this can be seen in Fig. \ref{fig:detailQSOS}, where regular and chaotic regions coexist: The zeros around the regular island seem to lie on the outmost invariant torus, while in the chaotic region they are scattered over the entire area.

\begin{figure*}
    \centering
    \includegraphics[width=0.92\linewidth]{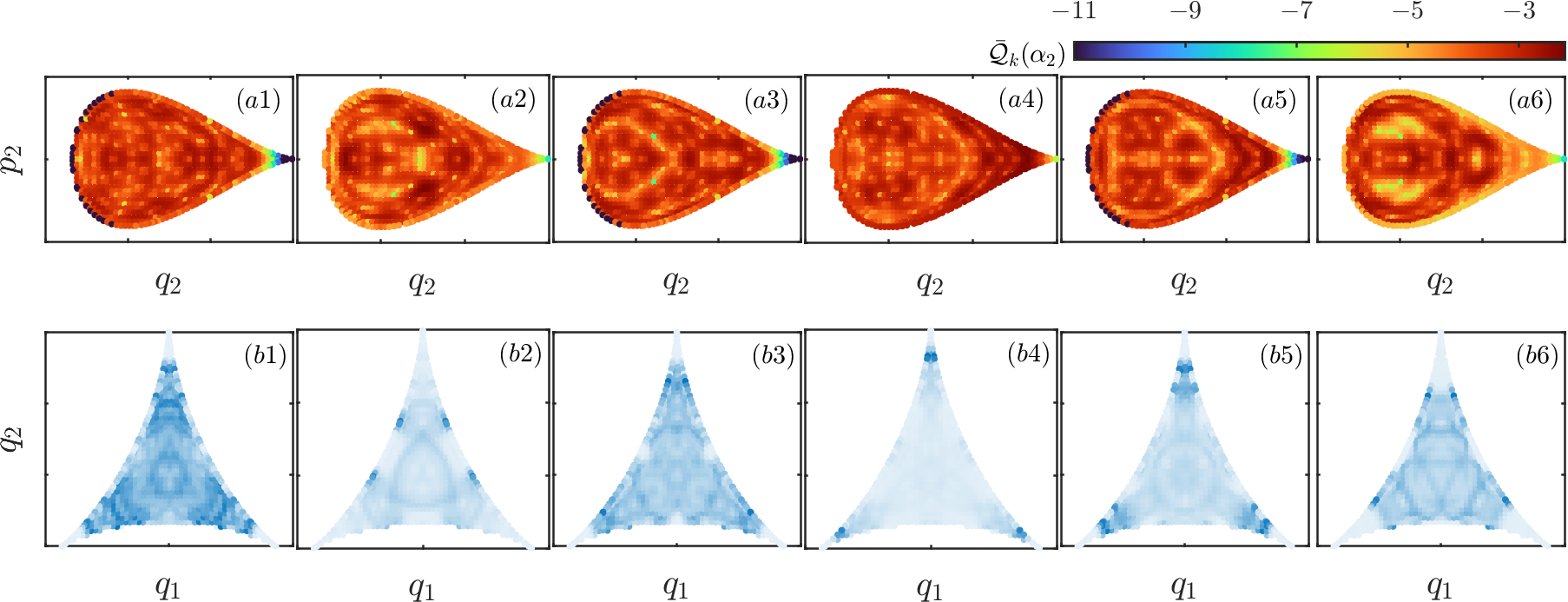}
    \caption{Examples of Husimi functions of six eigenstates in the energy interval $[E-\delta E/2,E+\delta E/2]$ at $E=1/3$ with $\delta E/E\simeq0.002$, for the singlet of  general FPUT with $\lambda=1/16$, where the cutoff of $n$ is set to $N=600$ and the Planck constant $\hbar=5\times 10^{-3}$. Panels ($a1$)-($a6$): Husimi QSOS given by Eq. \eqref{eq:QSOS-direct} plotted in the logarithmic scale. Panels $(b1)$-$(b6)$: projected Husimi functions in configuration space $\widetilde{\mathcal{P}}_k(q_1,q_2)$ given by Eq. \eqref{eq:husimiProjConfig} plotted in linear scale. Panels in the same column correspond to the same eigenstate.}
    \label{fig:genHus}
\end{figure*}

Moreover, we show in Fig. \ref{fig:genHus}($a1$)-($a6$) examples of the Husimi QSOS for the singlet of  general FPUT, with $\lambda=1/16$ and $\hbar=5\times 10^{-3}$, in an energy interval at $E=1/3$, the saddle energy (for more details of saddles, see paper I), with $\delta E/E\simeq0.002$ (in Appendix \ref{appD} there is a gallery of more quantum states). The classical SOS of the general case is shown in Fig. \ref{fig:HenonPSS}($c$),  and indicates that the system is classically fully chaotic at this energy, which has also been verified by our calculation of $\mu_c$, the relative phase space volume of the chaotic region (volume in the energy surface of the phase space by Liouville measure, see paper I).  All  Husimi QSOSs of the chaotic eigenstates are non-uniformly extended, and show different disordered pattern of weak zeros. The non-uniformity disappears in the semiclassical limit according to PUSC (see below).

\subsection{Projected Husimi function}
\label{sec3.2}
To identify quantum scars (or periodic orbits) embedded in  quantum states \cite{heller1984bound,berry1989quantum,gutzwiller2013chaos} and enhance the visualization of the Husimi function, we project the Husimi function  $\mathcal{H}_k(\alpha_1,\alpha_2)$ completely  into the phase space $(q_2,p_2)$ of the second oscillator,  as
\begin{align}
    \label{eq:completeProj}
    \mathcal{P}_k(\alpha_2)&=\frac{1}{\pi}\int \mathcal{H}_k(\alpha_1,\alpha_2)d^2\alpha_1\nonumber\\
    &=\sum_{n_1,n_2,n'_2}B_{n_1n_2}^kB_{n_1n'_2}^{k*}\langle \alpha_2|n_2\rangle\langle n'_2|\alpha_2\rangle,
\end{align}
by first integrate out the variable $\alpha_1$. $\mathcal{P}_k(\alpha_2)$ can be referred to as the \emph{completely} projected Husimi function, because it includes all energy shells, bringing about a blurred visualization of quantum scars \cite{pilatowsky2021quantum}.

Yet another definition of the projected Husimi function is by first intersecting the Husimi function with the classical energy shell at the respective eigenenergy $E_k$, and finalize the integration thereafter \cite{pilatowsky2022identification}, yielding
\begin{align}
    \label{eq:husimi-proj}
    \widetilde{\mathcal{P}}_k(\alpha_2)
&=\frac{1}{2\pi \hbar}\int \delta\left(E_k-H(\bf{q},\bf{p})\right)\mathcal{H}_k(\alpha_1,\alpha_2)dq_1dp_1\nonumber\\
    &=\frac{1}{2\hbar}\int_{p_1^-}^{p_1^+} dp_1 \sum_{q_1=q_1^\pm} \frac{\mathcal{H}_k(q_1,p_1;q_2,p_2)}{|\partial{H(\bf{q},\bf{p})}/\partial q_1|}\nonumber\\
    &=\frac{1}{\hbar}\int_{p_1^-}^{p_1^+} dp_1 \frac{\mathcal{H}_k(q_1^+,p_1;q_2,p_2)}{|\partial{H(\bf{q},\bf{p})}/\partial q_1|_{q_1=q_1^+}},
\end{align}
where $q_1^\pm$ are two roots of the equation $H(\textbf{q},\textbf{p})=E_k$, $p_1^\pm$ roots of the equation 
\begin{align}
    \frac{\partial{H}(q_1,p_1,q_2,p_2)}{\partial q_1}|_{q=q_1^\pm}=0,
\end{align}
and the Husimi function 
\begin{align}
    \mathcal{H}_k(q_1^+,p_1;q_2,p_2)=|\sum_{n,l}C_{nl}^k\langle \widetilde{\alpha}_+,\widetilde{\alpha}_-|n,l\rangle|^2,
\end{align}
where $\widetilde{\alpha}_1=(q_1^++ip_1)/\sqrt{2\hbar}$, $\widetilde{\alpha}_\pm =(\widetilde{\alpha}_1\mp i\alpha_2)/\sqrt{2}$, according to Eq. \eqref{eq:husimi-cart}.  It should be noted that the last integral for projected Husimi function can be computed effectively by a Chebyshev-Gauss quadrature method. In Fig. \ref{fig:HusimiEnsemble}$(b1)$-$(b6)$ we show the plots of  $\mathcal{P}_k(\alpha_2)$ and  in Fig. \ref{fig:HusimiEnsemble}$(c1)$-$(c6)$ the plots of $\widetilde{\mathcal{P}}_k(\alpha_2)$, for the same six consecutive eigenstates.  Comparing these two type of plots, it is evident that the completely projected Husimi function $\mathcal{P}_k(\alpha_2)$ is just the blurred projected Husimi function.  In all panels, the scaring of chaotic eigenstates is evident.

To further clarify the corresponding family of periodic orbits contributing to the scaring, it is much better to project the Husimi function into the configuration space. Similar to the method used to obtain the projected Husimi function in phase space of the subsystem, we can have the projected Husimi function in configuration space as 
\begin{align}
    \label{eq:husimiProjConfig}
    \widetilde{\mathcal{P}}_k(q_1,q_2)&=\frac{1}{2\pi \hbar}\int \delta\left(E_k-H(\bf{q},\bf{p})\right)\mathcal{H}_k(\alpha_1,\alpha_2)dp_1dp_2\nonumber\\
    &=\frac{1}{\hbar}\int_{p_2^-}^{p_2^+} dp_2\ \mathcal{H}_k(q_1,p_1^+;q_2,p_2)/p_1^+,
\end{align}
where $p_2^\pm=\pm\sqrt{2(E_k-V)}$, $p_1^+=\sqrt{2(E_k-V)-p_2^2}$, and $V(q_1,q_2)$ being the potential of the classical Hamiltonian. It is worth noting that the projected Husimi function $\widetilde{\mathcal{P}}_k(q_1,q_2)$ we have defined above, is related to the configuration-space probability density $|\psi_k(q_1,q_2)|^2$ of the eigenstate (see the derivation and flowchart shown in Fig. \ref{fig:flowchart}, Appendix \ref{appC}), where $\psi_k(q_1,q_2)$ can be written as a summation of the product of Hermite-Gauss modes 
\begin{align}
    &\psi_k(q_1,q_2)=\langle q_1,q_2|E_k\rangle =\sum_{n_1,n_2}B_{n_1n_2}^k\langle q_1,q_2|n_1,n_2\rangle\nonumber\\
    &=\sum_{n_1,n_2}B_{n_1n_2}^k\prod_{i=1}^2 \frac{(\hbar/\pi)^{-1/4}}{\sqrt{2^{n_i}n_i!}}H_{n_i}(\frac{q_i}{\sqrt{\hbar}})\exp(-\frac{q_i^2}{2\hbar}),
\end{align}
where $H_{n_i}(x)$ stands for the Hermite polynomial. Thus, for large $N$,  the numerical evaluation of $\psi_k(q_1,q_2)$ depends on the unavoidable direct calculation of large factorials, which make the computation much harder for $N>100$, while one can easily calculate the projected Husimi functions, even for $N>1000$.

In Fig. \ref{fig:HusimiEnsemble}($d1$)-($d6$) we present $\widetilde{\mathcal{P}}_k(q_1,q_2)$ the projected Husimi functions in configuration space for mixed-type system, of the same eigenstates which have been plotted in other panels using other Husimi functions.  As a comparison, we also show in Fig. \ref{fig:genHus}($b1$)-($b6$) $\widetilde{\mathcal{P}}_k(q_1,q_2)$ for the general FPUT, of which the corresponding classical dynamics is fully chaotic. The chaotic eigenstates in all projections, either for mixed-type system, or for fully chaotic system, display scarring that may be associated with one single particular periodic orbit from the corresponding classical dynamics, or as a superposition of several periodic orbits, similar to the case of Hydrogen atom in a uniform strong magnetic field \cite{robnik1981hydrogen,hasegawa1989classical,wintgen1989irregular}. In both $\alpha$-FPUT and the general case, because of the same $C_{3v}
$ symmetry of the potentials, we see the scars in Fig. \ref{fig:HusimiEnsemble}($d3$) and Fig. \ref{fig:genHus}($b2$), supported by the unstable periodic orbit from the same $B$ (``base'') family, which has been calculated by the  monodromy method \cite{davies1992calculations}. Other periodic orbits underlying different quantum scars, which have not been calculated by the monodromy method, are possible to be classified using the frequency analysis \cite{binney2011galactic,zotos2015classifying}, is subject of our forthcoming papers.

We close section \ref{sec3.2} with one of the main conclusions of this work: For a much more detailed study of the structure and the statistics of localization measure of chaotic eigenstates, one needs both Husimi QSOS and projected Husimi functions. The former is essential for the classification of different behaviors of eigenstates in mixed-type systems, because eigenstates in these systems can spread either over the integrable region, or a chaotic component, or there is the flooding of chaotic eigenstates from the chaotic region into the regular region, the so-called mixed states, if the semiclassical limit is not reached. The latter serves as a valuable tool for identifying highly localized states, and especially the periodic orbits underlying quantum scarred states.  These different types of Husimi functions work together to investigate the mechanisms of localization, enabling a comprehensive examination of this phenomenon.

\section{Fraction of mixed eigenstates}
\label{sec4}
Due to the correspondence between the Husimi QSOS and classical SOS, as shown in Sec. \ref{sec3.1}, we employ Husimi QSOS for the identification of regular, mixed, and chaotic eigenstates, by the criterion of overlap with the classical SOS, where one can use the SALI method to identify whether an initial condition on the classical SOS belongs to the chaotic or regular regions. As it was introduced and implemented in previous works, for the $k$-th eigenstate $|E_k\rangle$, the overlap is quantified by the overlap index $M_k$, defined as \cite{batistic2013dynamical,batistic2013intermediate,batistic2013quantum}
\begin{align}
    M_k=\int_{\mathcal{S}} dq_2dp_2 \ C(q_2,p_2) \bar{\mathcal{Q}}_k(\alpha_2),  
\end{align}
where $C(q_2,p_2)$ is $+1$ if  $(q_2,p_2)\in\mathcal{S}$ belongs to the chaotic region, and $-1$ if lies in the regular regions. By discretizing the classical SOS into a rectangular grid of points indexed by $(i,j)$ centered in cells of equal area, $M_k$ can be calculated numerically as 
\begin{align}
    M_k=\sum_{i,j}C_{i,j} \bar{\mathcal{Q}}_k^{i,j}/\sum_{i,j}\bar{\mathcal{Q}}_k^{i,j}, 
\end{align}
where $ C_{i,j}:=C(q_{2,i},p_{2,j})$, $\bar{\mathcal{Q}}_k^{i,j}:=\bar{\mathcal{Q}}_k(q_{2,i},p_{2,j})$. 

\begin{figure}
    \includegraphics[width=1\linewidth]{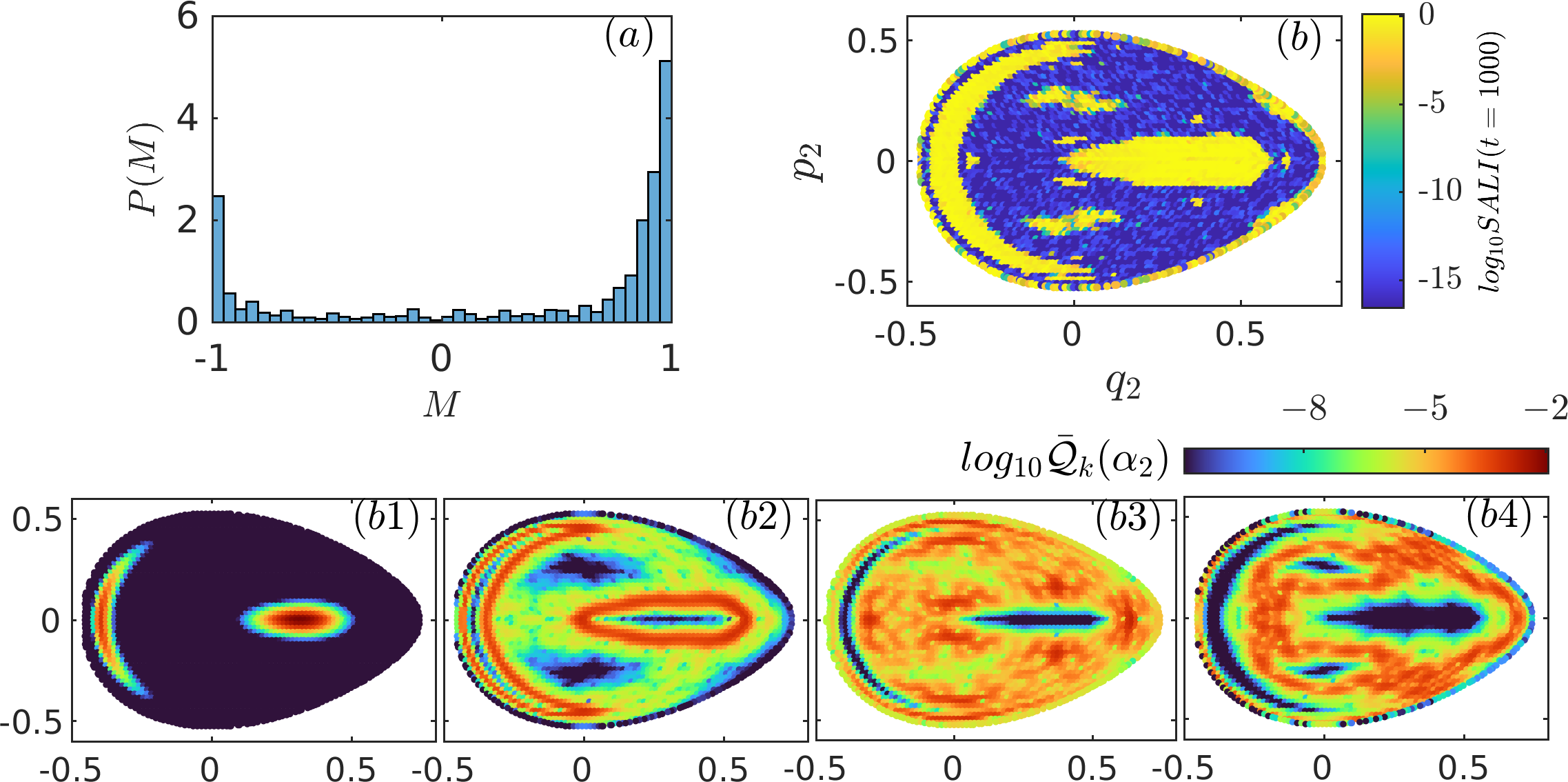}
    \caption{($a$) The histogram of distribution of the overlap index $M$ for an ensemble of approximately 1200 eigenstates, in the energy interval $[E-\delta E/2,E+\delta E/2]$ at $E=0.14$ with $\delta E/E\simeq0.01$, for the singlet of $\alpha$-FPUT,  where the irreducible Hilbert space is of size $\mathcal{N}_S\sim 3.75\times10^5$, with the cutoff $N=1500$ and the Planck constant $\hbar=(4\pm0.016)\times 10^{-4}$. ($b$) Regions of different (logarithmic) values of the SALI on the classical SOS for $E=0.14$ the same as  Fig. \ref{fig:HenonPSS}(a), at $t=1000$: the initial conditions colored dark blue correspond to chaotic orbits, the yellowish color indicates ordered motion, and the intermediate color suggests sticky orbits. Panels ($b1$)-($b4$): Husimi QSOS plotted in the logarithmic scale, for four eigenstates selected from the ensemble, with different values of $M$ index,  from left to right $M\simeq -1,-0.5,0.5,1$ respectively, where the darkest blue indicates area where $\bar{\mathcal{Q}}_k(\alpha_2)< 10^{-11}$. }
    \label{fig:mixedHenonMHusimi}
\end{figure}

According to PUSC of Husimi functions, in the sufficiently deep semiclassical limit, $M_k$ should be either $+1$ or $-1$, for chaotic or regular Husimi QSOS, respectively. However, as it has been revealed in systems such as quantum billiards, the actual value of $M_k$ varies from $-1$ to $+1$, if the semiclassical limit is not yet reached. This is confirmed by the histogram shown in Fig. \ref{fig:mixedHenonMHusimi} ($a$), where we show the distribution of the overlap index $M$ for an ensemble of approximately 1200 eigenstates, in the energy interval $[E-\delta E/2,E+\delta E/2]$ at $E=0.14$ with $\delta E/E\simeq0.01$, for the singlet of $\alpha$-FPUT. It exhibits a bimodal distribution, where the two peaks are located at each end with $M=-1$ and $M=+1$. In the calculation of $M$, we use the SALI method to numerically compute $C_{i.j}$ across the SOS. The criterion for the classification of initial conditions belonging to a chaotic region is that SALI $\le 10^{-8}$ at t = 1000 (the unit of dimensionless time is one period of the linear oscillator, for more details about this criterion, see paper I). Fig. \ref{fig:mixedHenonMHusimi}($b$) shows the SALI plot on the classical SOS, where the value of SALI of each point is plotted using an assigned color accordingly. It shows some similarity compared with classical SOS shown in Fig. \ref{fig:HenonPSS}$(a)$, but there are more details visible especially the stickiness on the border between the regular and chaotic region. The relative area of the chaotic components on the classical SOS $\eta_c$ computed by SALI is 0.674 at $E=0.14$, where 
\begin{align}
    \eta_c = \frac{\int_{\mathcal{S}} dq_2dp_2 \chi_c(\textbf{q},\textbf{p})}{A(\mathcal{S})},
\end{align}
where $\text{A}(\mathcal{S})$ is the surface area of $\mathcal{S}$ the classical SOS, $\chi_c(\textbf{q},\textbf{p})$ denotes the characteristic function of the chaotic component, which takes the value of 1 on chaotic region and zero otherwise. 

\begin{figure}
    \includegraphics[width=1\linewidth]{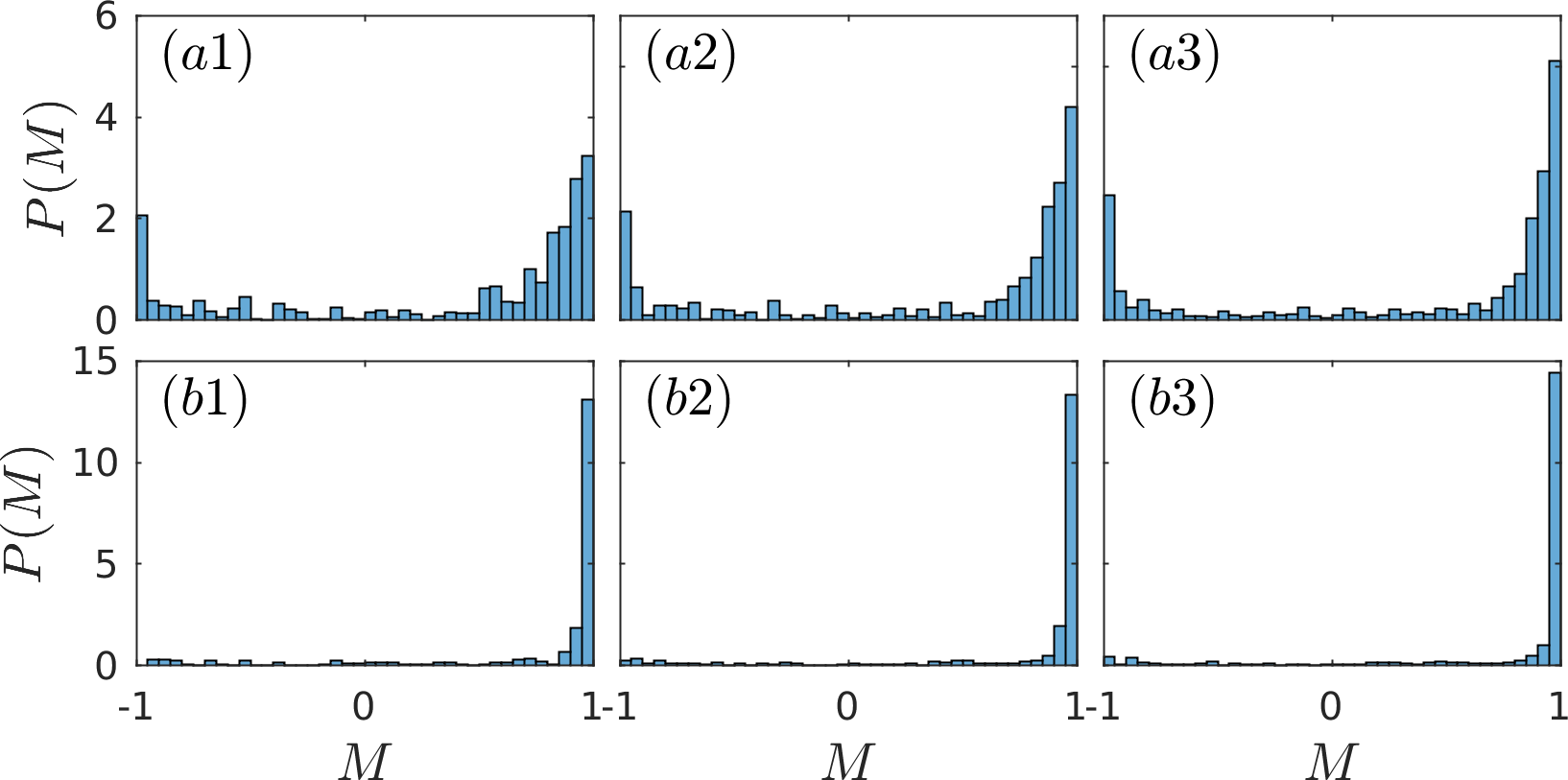}
    \caption{The histogram of distribution of overlap index $M$ for an ensemble of approximately 1200 eigenstates from the singlet of $\alpha$-FPUT, in the energy interval $[E-\delta E/2,E+\delta E/2]$ with $\delta E/E\simeq0.01$, for energies $E=0.14$ (top panels), and $E=0.16$ (bottom panels). The Planck constant is $\hbar=\hbar_0 \pm j\hbar_0/1000$, where from left to right  $\hbar_0=1\times 10^{-3} (\text{with} \ j=0,1,2,3,4,5), 7\times 10^{-4} (\text{with} \ j=0,2,4), 4\times 10^{-4}(\text{with} \ j=4)$.}
    \label{fig:mixedHenonMHist}
\end{figure}

In Fig. \ref{fig:mixedHenonMHusimi}($b1$)-($b4$) we show four typical examples of Husimi QSOS from the ensemble,  with different values of $M$. Except the $M=-1$ state living in the regular islands and the $M=+1$ state extending across the chaotic region, there are mixed eigenstates with $|M|<1$: the $M\simeq -0.5$ state shown in Fig. \ref{fig:mixedHenonMHusimi}$(b2)$ lives predominantly in the vicinity of the regular islands with a small contribution in the main part of the chaotic sea, which is referred to as the hierarchical state in the study of quantum standard map \cite{ketzmerick2000new}. It also exhibits the tunneling process between the different islands. Most importantly, there is the $M\simeq 0.5$ state shown in Fig. \ref{fig:mixedHenonMHusimi}$(b3)$ that is flooding from the chaotic sea to the regular region. While, it must be pointed out, the state shown in Fig. \ref{fig:HusimiEnsemble}$(a6)$ is of $M\simeq 0.52$, although predominantly inside the chaotic region, it is much less extended, due to dynamical localization. This reveals that the overlap index $M$ is not directly correlated with the extendedness, which we will discuss in more detail in Sec. \ref{sec5}, studying the statistics of the localization measures. 

To illustrate a general trend in the reduction of mixed eigenstates as one approaches a deeper semiclassical limit, we have calculated the $M$ index of an ensemble of about 1200 eigenstates, in narrow energy intervals with $\delta E/E\simeq0.01$,  for different values of the Planck constant. It is worth emphasizing here in the numerical calculation of Husimi QSOSs, there is an unavoidable computing limitation of the smallness of $\hbar$, due to the exponential term $e^{-|\alpha|^2/2} = e^{-(p^2+q^2)/2\hbar}$ in the definition of coherent state basis. From the analytical expression of the semiclassical density of states, consequently there is a limitation of the number of energy levels in the energy intervals. To achieve a statistically reasonable size of each ensemble, we set $\hbar=\hbar_0\pm j\hbar_0/1000$, where the chosen $j$ ensures the $\hbar\in[\hbar_0-\delta \hbar_0/2,\hbar_0+\delta \hbar_0/2]$ with $\delta \hbar_0/\hbar_0 \le 0.01$. The narrow energy window and narrow window of Planck constant ensure that eigenstates from the same ensemble have approximately the same classical correspondence. Fig. \ref{fig:mixedHenonMHist} shows the distribution of $M$ index at two energies, for three different $\hbar$.  The relative area of chaotic components is $\eta_c=0.865$ on classical SOS at $E=0.16$, therefore the relative fraction of $M=1$ states shown in Fig. \ref{fig:mixedHenonMHist}$(b)$ is larger than in the case $E=0.14$ shown in Fig. \ref{fig:mixedHenonMHist}$(a)$, where $\eta_c=0.674$, and for the $M=-1$ states the opposite.

\begin{figure}
    \includegraphics[width=0.95\linewidth]{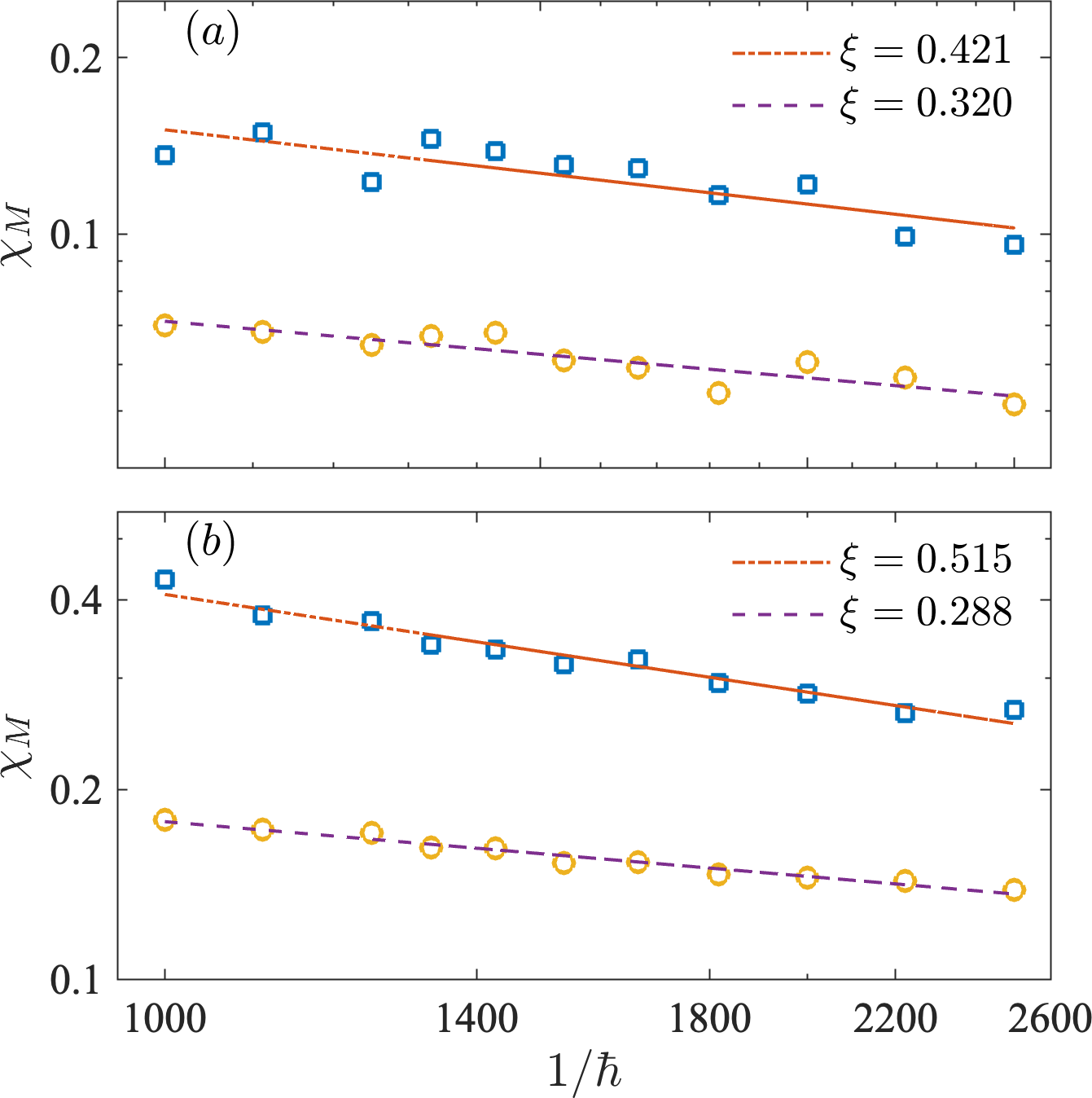}
    \caption{Decay of the fraction of mixed eigenstates $\chi_M(E,\hbar)$ with respect to the Planck constant $\hbar$, at two energies $E=0.14$ (squares) and $E=0.16$ (circles), where $\delta E/E \simeq 0.01$ and $N(E,\delta E, \hbar)\approx 1200$. The dash-dotted lines and dashed lines show the power law $\chi_M(E,\hbar) \sim \hbar^{\xi}$, where $(a)$ is for mixed eigenstates with $-0.8\le M\le 0$, and in $(b)$  $-0.8\le M\le 0.8$.}
    \label{fig:mixedPowerDecay}
\end{figure}

With the decreasing $\hbar$, the relative fraction of $M=\pm 1$ states increases accordingly, meaning that increasingly more states either live in the purely regular region or the chaotic sea, while the fraction of mixed eigenstates decreases, when approaching a deeper semiclassical limit. To quantify the decay of the fraction of mixed states, we define 
\begin{align}
    \chi_M(E,\hbar)=\frac{N(M\in\{M_0,M_1\}\ |E,\delta E,\hbar)}{N(E,\delta E,\hbar)}
\end{align}
as the fraction of eigenstates with the overlap index $M_0\le M\le M_1$, given  the total number of states $N(E,\delta E, \hbar)$ in an energy interval, within $\delta E\ll E$. For the case $-0.8 \le M \le 0$, the mixed states with more regular region occupation, we see in Fig. \ref{fig:mixedPowerDecay}$(a)$ a power-law decay of $\chi_M(E,\hbar) \sim \hbar^{\xi}$ with respect to the decreasing value of $\hbar$, where $\xi=0.421$ for $E=0.14$ and $\xi=0.320$ for $E=0.16$. For mixed eigenstates with $|M|\le 0.8$, with states neither predominantly in the regular islands nor in the chaotic sea, $\chi_M(E,\hbar)$ also follows a power law of $\hbar$, but with slightly different power exponents, as shown in Fig. \ref{fig:mixedPowerDecay}$(b)$. It should be noted that the power-law decay of the fraction of mixed states we have found here is different from the result in Ref. \cite{ketzmerick2000new} for quantum standard map, regarding the power-law decay of the hierarchical states, which generally predominantly live in the vicinity of the regular islands, by definition the $M\le 0$ states.

Upon examining various power-law decays in Fig. \ref{fig:mixedPowerDecay},  it was observed that the states with $M\le0$ exhibit a slower decay rate than those with $M>0$ at  $E=0.14$. Conversely, at $E=0.16$, the trend reverses, with the $M>0$ states demonstrating a slower decay compared to the $M\le0$ states. While we have shown in Sec. \ref{sec3.1} and Fig. \ref{fig:henonGallery} in Appendix \ref{appD}, for states with $M \leq 0$, the predominant mechanism involves (chaos-assisted) tunneling between regular islands, whereas for states with $M>0$, the primary phenomenon is the flooding of chaotic states into regular regions. It delineates the disparities between these two tunneling mechanisms as one approaches the semiclassical limit.

\section{Localization measures of chaotic eigenstates}
\label{sec5}
The study of Husimi functions in Sec. \ref{sec3}, either the Husimi QSOS or the projected Husimi functions, clearly demonstrates the localization of chaotic eigenstates to varying degrees within the same narrow energy interval, in both $\alpha$-FPUT the mixed-type system and the general case with $\lambda=1/16$ which is fully chaotic, ergodic. To further quantify the extent of localization, following the developed method that has been used to reveal the scaling behavior of localization \cite{izrailev1988quantum,casati1990scaling} in natural basis, and recently for the study of  Poincar\'e-Husimi functions in different type of quantum billiards (see Ref. \cite{batistic2019statistical,lozej2022quantum}), we define localization measures based on the R\'enyi-Wehrl entropy \cite{gnutzmann2001renyi} of the Husimi QSOS. While the entropy of Husimi QSOS of the chaotic eigenstate $|E_k\rangle $ is given as
\begin{align}
    \label{eq:rw1}
  H_k^{(\alpha)}:= \frac{1}{1-\alpha}\ln  \int_{\mathcal{S}_c}dq_2dp_2 \left[\mathcal{Q}_k(\alpha_2)\right]^{\alpha},
\end{align}
where $(q_2,p_2)\in \mathcal{S}_c$ and $\mathcal{S}_c$ (compact phase space) denotes the chaotic region on the classical SOS, $\mathcal{Q}_k(\alpha_2)$ is the rescaled Husimi QSOS
\begin{align}
    \label{eq:rw2}
    \mathcal{Q}_k(\alpha_2)= \bar{\mathcal{Q}}_k(\alpha_2)/\int_{\mathcal{S}_c} dq_2dp_2 \bar{\mathcal{Q}}_k(\alpha_2).
\end{align}
The R\'enyi-Wehrl entropy localization measure (ELM) is then defined as 
\begin{align}
    \label{eq:elm}
    \mathcal{L}_k^\alpha = \frac{\exp H_k^{(\alpha)}}{\text{A}(\mathcal{S}_c)},
\end{align}
where $\text{A}(\mathcal{S}_c)$ is the surface area of $\mathcal{S}_c$. Obviously, for fully chaotic systems $\mathcal{S}_c\equiv\mathcal{S}$ and $\mathcal{Q}_k(\alpha_2)\equiv\bar{\mathcal{Q}}_k(\alpha_2)$. The latter equivalence is also true for mixed-type systems, if it is in the semiclassical limit. But as we have shown in Sec. \ref{sec4}, for a regime that is not deeply semiclassical, there exist states that live both in the regular and chaotic region, with the overlap index $|M|<1$, thus $\mathcal{Q}_k(\alpha_2)\ne \bar{\mathcal{Q}}_k(\alpha_2)$, and it becomes necessary to establish a criterion for identifying chaotic eigenstates: states with $M\ge M_c$ can be declared chaotic.  Ref. \cite{batistic2013dynamical} has demonstrated that there exist two natural criteria, one from classical referring to the relative Liouville measure of the classical part of the phase space $\mu_c$, the other from quantum, referring to the agreement of the chaotic level statistics with the Brody distribution. In the following, we choose $M_c=0.8$, so that the states predominantly live in the chaotic region, to just make sure that only chaotic eigenstates are being used.

In the limit $\alpha\to 1$, Eq. \eqref{eq:rw1} gives the Wehrl entropy, an information entropy based on Husimi functions. The $\alpha=2$  R\'enyi-Wehrl ELM is the inverse participation ratio (IPR). The R\'eyni-Wehrl entropy is a non-increasing function of the parameter $\alpha$, and as $\alpha$ tends towards infinity, it is increasingly determined by events of highest probability, therefore quite straightforwardly, one can use higher order R\'enyi-Wehrl ELM to detect highly localized or scarred states, as demonstrated in recent studies of the Dicke model with projected Husimi functions  (see Ref. \cite{pilatowsky2022identification}). There are asymptotic upper bounds $\mathcal{L}_{|\mathcal{R}\rangle}^{\alpha}$ for the R\'enyi-Wehrl ELMs, from the eigenvector statistics of random unitary matrices, where $|\mathcal{R}\rangle$ is a random pure state of size $N\gg 1$. First, it leads to an ensemble average of the R\'enyi-Wehrl entropy \cite{jones1990entropy,zyczkowski1994random}
\begin{align}
    \langle H_{|\mathcal{R}\rangle}^{(\alpha)}\rangle_N=\frac{1}{1-\alpha}\ln\frac{\Gamma(N+1)\Gamma(\alpha+1)}{\Gamma(N+\alpha)}.
\end{align}
 We can then define the corresponding R\'enyi-Wehrl ELM in large $N$ limit as the upper bounds
\begin{align}
    \mathcal{L}_{|\mathcal{R}\rangle}^{\alpha}=\lim_{N\to\infty} \frac{\exp\langle H_{|\mathcal{R}\rangle}^{(\alpha)}\rangle_N}{N}=\Gamma(1+\alpha)^{1/(1-\alpha)}.
\end{align}
Then, one has 
\begin{align}
     \mathcal{L}_{|\mathcal{R}\rangle}^{1}= e^{\gamma-1}, \quad 
     \mathcal{L}_{|\mathcal{R}\rangle}^{2}= 0.5,
\end{align}
where $\gamma$ denotes the Euler constant so that $\mathcal{L}_{|\mathcal{R}\rangle}^{1}\approx 0.66$. It is necessary to point out that these upper bounds do not contradict the microcanonical assumption \cite{voros1977asymptotic} for an ergodic system and the PUSC for the mixed-type system, namely that in the semiclassical limit, the \emph{local} averaged Wigner (or Husimi) function can be written as
\begin{align}
    \bar{\Psi}_k(\textbf{q},\textbf{p})=\frac{\delta(E_k-H(\textbf{q},\textbf{p}))\chi_c(\textbf{q},\textbf{p})}{\int d\textbf{q} d\textbf{p}\delta(E_k-H(\textbf{q},\textbf{p}))\chi_c(\textbf{q},\textbf{p})},
\end{align}
 which means that the state is uniformly extended over the entire phase space in the fully chaotic system, and over chaotic invariant components in the mixed-type. This uniformity results in $\alpha$ R\'enyi-Wehrl ELM of the averaged state to be $\mathcal{L}^\alpha_k =1$. The ``\emph{local}" means that the average is taken over many oscillations of the wavefunction, while the scale of these oscillations is of order $\hbar$, as pointed out by Berry (see Ref. \cite{berry1977regular}).

Thus, following the statistics given by random matrix theory (RMT), the localization measure of the (fully extended) chaotic state is $\mathcal{L}_k^\alpha=\mathcal{L}_{|\mathcal{R}\rangle}^\alpha$, calculated from Husimi functions without local average.  The localization phenomena in the chaotic eigenstates implying $\mathcal{L}^\alpha_k < \mathcal{L}^\alpha_{|\mathcal{R}\rangle}$, are determined by the ratio of the Heisenberg time $t_H$ and the classical transport time $t_T$, as the localization control parameter, denoted as
\begin{align}
    \alpha_\mathcal{L} = t_H/t_T,
\end{align}
where $t_H=2\pi \hbar\rho(E)$ with $\rho(E)$ the quantum density of states (DOS). The chaotic eigenstates are maximally localized if  $\alpha_\mathcal{L}\ll 1$, and maximally extended if $ \alpha_\mathcal{L} \gg 1$, in between there is a distribution of the localization measures. 

\subsection{Classical transport time and Heisenberg time}

\begin{figure}
    \includegraphics[width=1\linewidth]{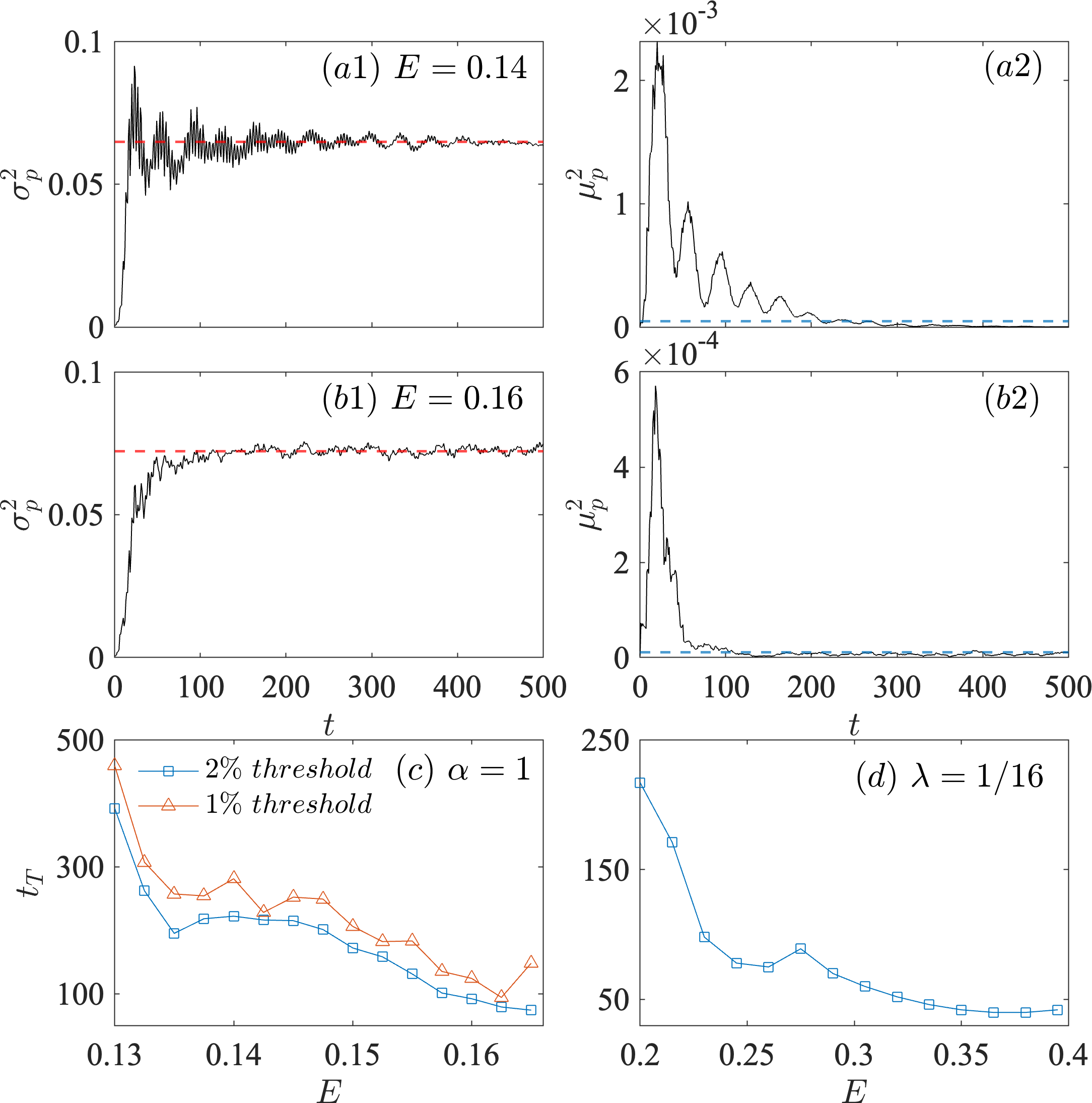}
    \caption{$(a1)$ The variance of momentum $\sigma_p^2=\langle p^2\rangle-\langle p\rangle^2$ vs. $t$ for $\alpha$-FPUT with $\alpha=1$ at $E=0.14$, where $\sigma_p^2$ is calculated with 4000 initial conditions from the classical SOS that are uniformly distributed in the chaotic region with $q_2\in[-0.25,-0.15]$ and $p_2=0$. The red dashed horizontal line indicates  $\overline{\sigma_p^2}$, the longtime averaged value of  $\sigma_p^2$. $(a2)$ The temporal fluctuation of $\sigma_p^2$, as a function of time,  across the time windows $[t-\Delta t/2,t+\Delta t/2]$, with $\Delta t=30$, and the blue dashed horizontal  line indicates the threshold $2\%$ of the peak value of $\mu_p^2$. $(b1)$-$(b2)$ Analogous data as $(a1)$-$(a2)$ respectively, at $E=0.16$. $(c)$ Classical transport time $t_T$ as a function of energy $E$ in $\alpha$-FPUT (H\'enon-Heiles), using two different thresholds. $(d)$ $t_T$ versus  $E$ in general FPUT ($2\%$ threshold).}
    \label{fig:ctransport}
\end{figure}

The classical transport (diffusion) time $t_T$ as a classical timescale, is the typical time needed for an ensemble of sharply distributed initial momentum with zero variance, to spread uniformly over the classical chaotic component. That is, one needs to estimate the timescale $t_T$ of the momentum spreading characterized by $\sigma^2_p(t)$, the variance of momentum, at which it reaches a certain saturation value. While the saturation value can be obtained from the longtime average $\overline{\sigma_p^2}$, the estimation of the exact time of reaching it is not obvious, as shown in Fig. \ref{fig:ctransport}($a1$)-($b1$), because in practice there are fluctuations around $\overline{\sigma_p^2}$. As an analog to the definition of quantum relaxation time \cite{cramer2008exact,cramer2010quantum}, we define the classical transport time as
\begin{align}
    \label{eq:ClassicalTransport}
   | \sigma^2_p(t)- \overline{\sigma_p^2}| \le \epsilon_1, \text{\ for all\ } t_T\le t \le t_{rec},
\end{align}
where $\epsilon_1$ is some small quantity and $t\le t_{rec}$ is the classical recurrence time. Another equivalent definition is that $\mu_p^2(t)$ the temporal fluctuation of $\sigma_p^2(t)$ across time windows $[t-\Delta t/2,t+\Delta t/2]$, as shown in Fig. \ref{fig:ctransport}$(a2)$-$(b2)$, fulfills that $\mu_p^2 \le \epsilon_2$ for $t\ge t_T$, $\epsilon_2$ as the threshold is also a small quantity.

Comparing these two definitions and the correspondent numerical results shown in Fig. \ref{fig:ctransport}, we found that practically it is more convenient to estimate $t_T$ according to the temporal fluctuation. In Fig. \ref{fig:ctransport}(c) we show $t_T$ as a function of energy $E$ in $\alpha$-FPUT, with two different thresholds: $1\%$ and $2\%$ of the peak value of $\mu_p^2(t)$. The general behavior of the transport time versus energy is similar under two thresholds, as expected, and it exhibits an almost monotonic decrease as energy increases, because chaos is more pronounced (larger $\mu_c$) at higher energies within this range of energy. In the following, we opt for the $2\%$ threshold in our estimation for two specific reasons:  it aligns more consistently with the results obtained from the definition in Eq. \eqref{eq:ClassicalTransport} and exhibits fewer anomalies compared to the $1\%$ threshold. For the general FPUT with $\lambda=1/16$, numerical results of $t_T$ are shown in Fig. \ref{fig:ctransport}$(d)$.

\begin{figure}
    \includegraphics[width=0.95\linewidth]{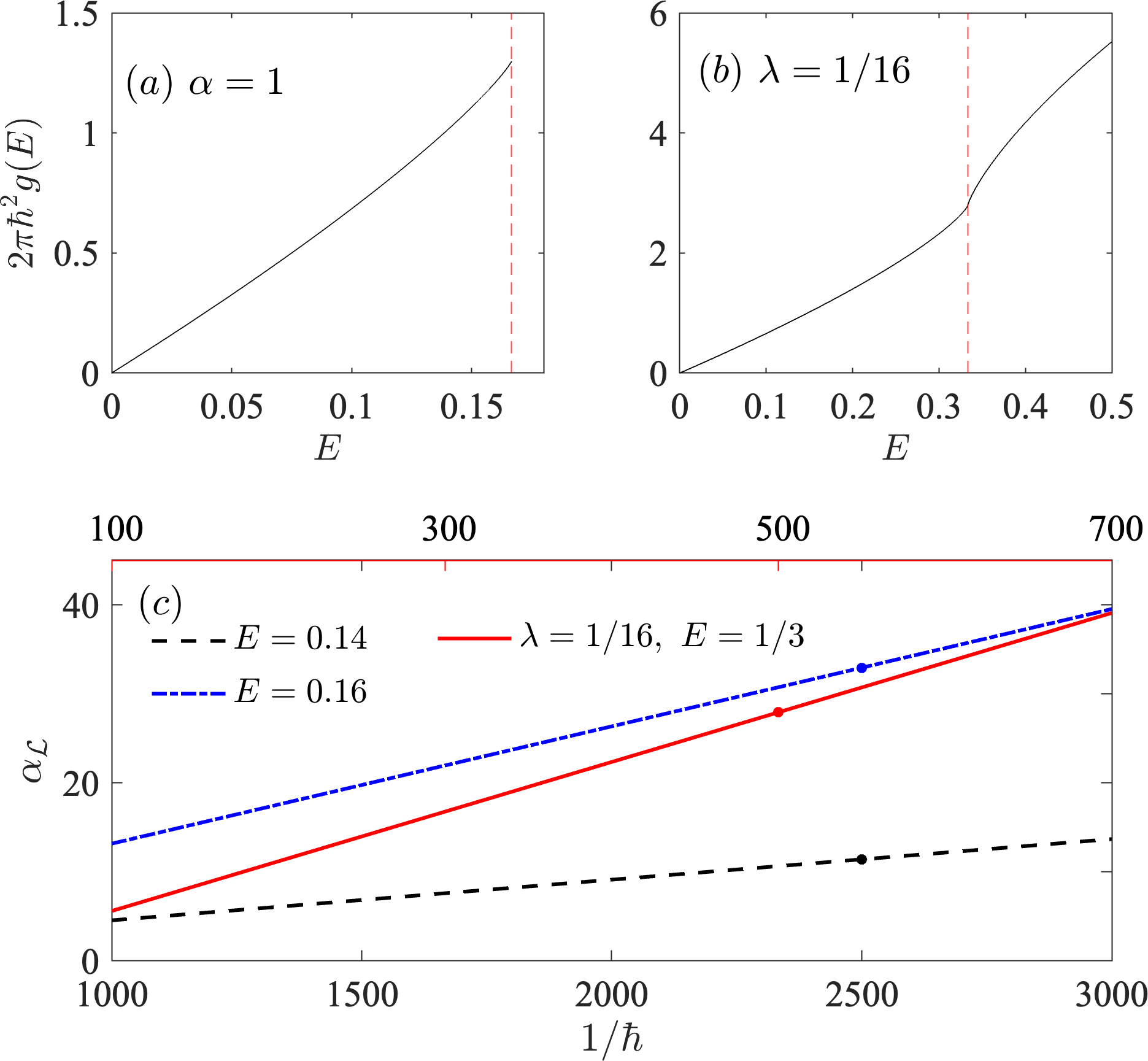}
    \caption{Scaled semiclassical DOS  $2\pi\hbar^2g(E)$ for $\alpha$-FPUT $(a)$ and the general case $(b)$, where the red dashed vertical lines denote the saddle energies (see paper I).  $(c)$ Ratio $\alpha_\mathcal{L}$ as a function of $1/\hbar$,  where black dashed line is for $E=0.14$ and blue dash-dotted line for $E=0.16$, in $\alpha$-FPUT, with respect to the bottom x-axis (colored in black). On the lines the black and blue dot indicate $\alpha_\mathcal{L}$ at $\hbar=4\times 10^{-4}$. The red solid line is for $E=1/3$ in general FPUT, with respect to the top x-axis in red color, where the red dot indicates the ratio at $\hbar=2\times 10^{-3}$.}
    \label{fig:alphaLocalization}
\end{figure}

Regarding the Heisenberg time $t_H= 2\pi \hbar\rho(E)$, the quantum DOS $\rho(E)$ has a semiclassical approximation $g(E)$, obtained from the Thomas-Fermi rule
  \begin{align}
      g(E)=\frac{1}{(2\pi\hbar)^2}\int d\textbf{q}d\textbf{p}\delta(E-H(\textbf{q},\textbf{p})),
  \end{align}
  which is the volume of the available classical phase space for a given energy $E$ divided by $(2\pi\hbar)^2$. For $\alpha$-FPUT of the H\'enon-Heiles potential, when $E\le 1/6\alpha^2$, the resulting Thomas-Fermi expression \cite{yan2024chaosPaperI} is 
\begin{align}
  g(E) =\frac{3}{4\pi\hbar^2}\big[\int_0^{\pi/3}r_+^2d\phi+\int^{2\pi/3}_{\pi/3}r_-^2d\phi\big],
\end{align}
where
\begin{align}
  \begin{cases}
  r_+(\phi) = \frac{1}{\alpha\sin 3\phi}(\cos\frac{\theta}{3}-\frac{1}{2}), &\text{if } \sin3\phi \ge0, \\
  r_-(\phi)=\frac{-1}{\alpha\sin 3\phi}(\cos\frac{\theta+\pi}{3}+\frac{1}{2}), &\text{if }\sin3\phi \le0,
  \end{cases}
\end{align}
and $\cos\theta = 12E\alpha^2\sin^23\phi-1$, $\theta\in[0,\pi]$. For the general case with $\lambda=1/16$, an analytical expression of $g(E)$ depends on the solving of a quartic equation (for a detailed derivation, see paper I). The scaled semiclassical DOS $f(E)=2\pi\hbar^2g(E)$ for two cases, as a function of energy, is shown in Fig. \ref{fig:alphaLocalization}$(a)$-$(b)$. Therefore, the controlling ratio
\begin{align}
    \alpha_{\mathcal{L}}=\frac{2\pi\hbar\rho(E)}{t_T}\simeq\frac{2\pi\hbar g(E)}{t_T} =\frac{f(E)}{\hbar t_T}.
\end{align}

In Fig. \ref{fig:alphaLocalization}$(c)$ we show $\alpha_{\mathcal{L}}$ as a function of $1/\hbar$, at two energies for $\alpha$-FPUT, and at the saddle energy $E=1/3$ for the general case. The increase in $\alpha_{\mathcal{L}}$ in relation to $1/\hbar$ at $E=0.14$ is slower compared to $E=0.16$,  attributable to the larger classical transport time at the former energy. While, in the general case at $E=1/3$ with $\hbar=2\times 10^{-3}$, the ratio is comparable to that of the $\alpha$-FPUT system at $E=0.16$ with $\hbar=4\times 10^{-4}$. These distinctions and comparisons will manifest in the distribution of localization measures, as we will illustrate in the subsequent discussion.

\begin{figure*}
    \includegraphics[width=0.95\linewidth]{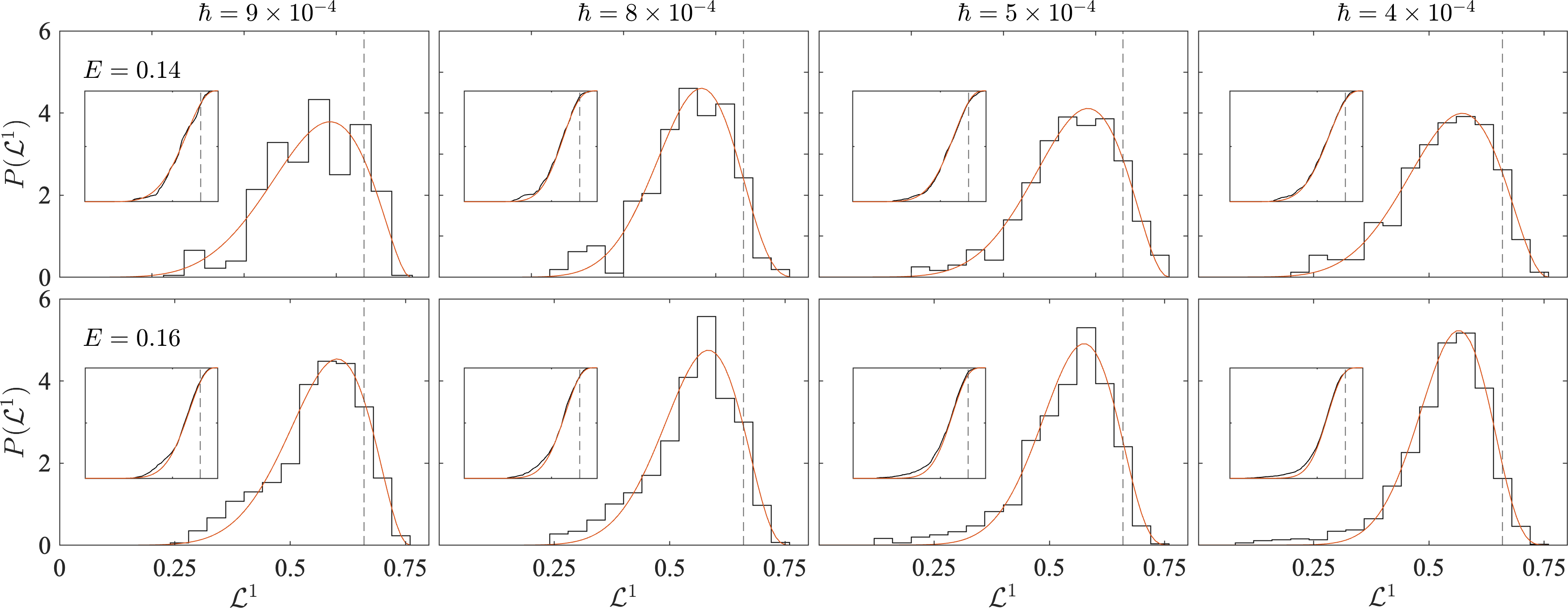}
    \includegraphics[width=0.95\linewidth]{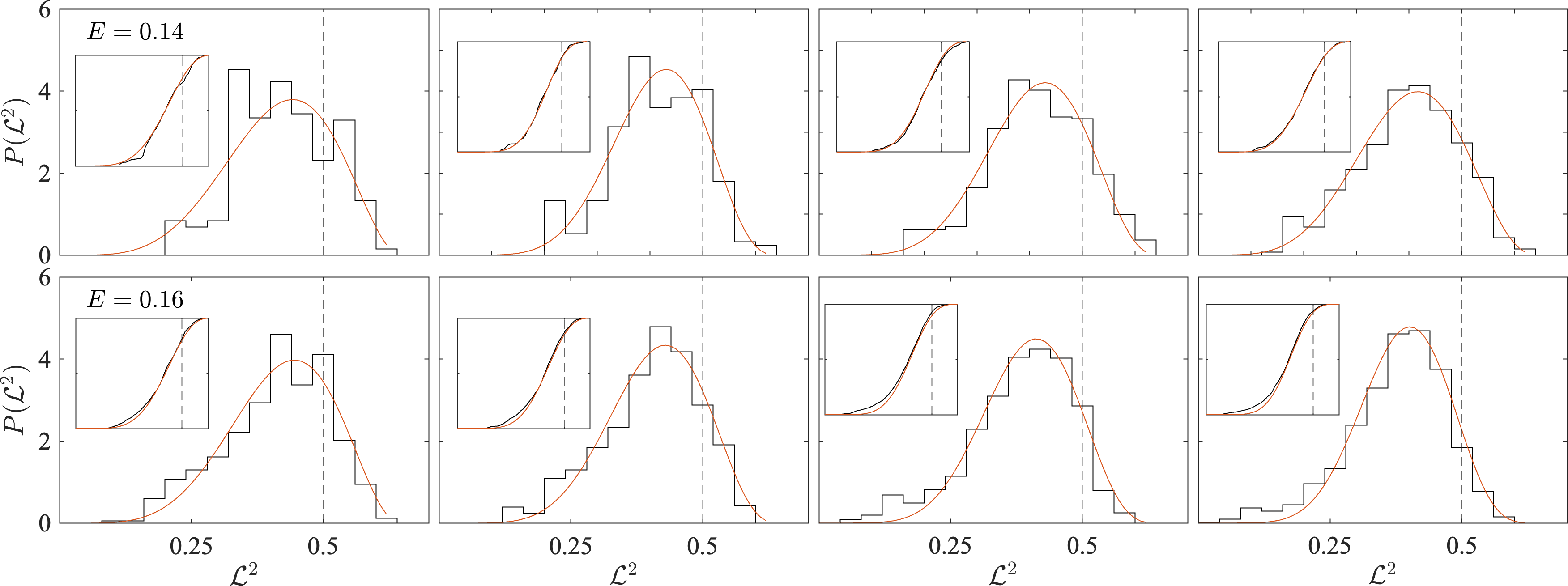}
    \caption{Distributions of the $\alpha=1$ and $\alpha=2$ R\'enyi-Wehrl ELMs, for the $M\ge 0.8$ eigenstates, as a function of the Planck constant $\hbar$ (indicated at the top of each column), at two energies for the singlet of $\alpha$-FPUT. The main figures show the histogram and its best-fitting beta distribution, while the insets show the corresponding cumulative distribution. (Colors online: black are data, red is the best-fitting). The vertical dash lines indicate $\mathcal{L}^1_{|\mathcal{R}\rangle}\approx 0.66$ for $\alpha=1$ cases and $\mathcal{L}^2_{|\mathcal{R}\rangle}= 0.5$ for $\alpha=2$, from random pure states. Parameters of the best fit beta distribution for  $P(\mathcal{L}^1)$ from left to right: at $E=0.14$ are $(6.17,1.83)$, $(10.72,3.98)$, $(7.48,2.26)$, $(7.18,2.35)$ with $\mathcal{L}^1_{0}= 0.76$ and at $E=0.16$ are $(8.87,2.33)$, $(10.29,3.11)$, $(11.27,3.62)$, $(13.17,4.55)$ respectively with $\mathcal{L}^2_{0}= 0.64$. Eigenstates for the statistics in each panel, are selected from an ensemble of $N(E,\delta E, \hbar)\approx 1200$ states with $\delta E/E \simeq 0.01$. }
    \label{fig:localizationHenon}
\end{figure*}

\subsection{The distribution of localization measures}
The distributions of R\'enyi-Wehrl ELMs have been studied thoroughly in different families of billiards (with or without stickiness, mixed or ergodic), demonstrating that the ELMs of a sequence of consecutive eigenstates conform to a shared empirical distribution, well described by the beta distribution \cite{batistic2019statistical,batistic2020distribution}, namely
\begin{align}
    P(\mathcal{L}^\alpha)=\frac{1}{C}(\mathcal{L}^\alpha)^{\beta_a}(\mathcal{L}_0^\alpha-\mathcal{L}^\alpha)^{\beta_b},
\end{align}
where $\mathcal{L}^\alpha_0$ is the upper limit of the interval $[0,\mathcal{L}^\alpha_0]$ on which $P(\mathcal{L}^\alpha)$ is defined, $\beta_a$ and $\beta_b$ are two positive shape parameters, the normalization constant $C$ is given by 
\begin{align}
    C=(\mathcal{L}_0^\alpha)^{\beta_a+\beta_b+1}B(\beta_a+1,\beta_b+1),
\end{align}
where $B(x,y)=\int_0^1 t^{x-1}(1-t)^{y-1}dt$ is the beta function. This has also been confirmed in the Dicke model and in the kicked-top model \cite{wang2020statistical,wang2023statistics}. Given the variance of beta distribution 
\begin{align}
    \sigma^2(\mathcal{L}^\alpha) = (\mathcal{L}^\alpha_0)^2\frac{(\beta_a+2)(\beta_b+2)}{(\beta_a+\beta_b+4)(\beta_a+\beta_b+3)^2},
\end{align}
therefore the variation in shape parameters is directly associated with the degree of localization of the eigenstates, controlled by the ratio $\alpha_\mathcal{L}$: A greater value of the parameter $\beta_a$ generally results in a more compact distribution with a reduced width, and in the limit $\beta_a\to \infty$, $P(\mathcal{L}^\alpha)=\delta(\mathcal{L}^\alpha-\mathcal{L}^\alpha_0)$, while the statistics from RMT shows that $P(\mathcal{L}^\alpha)=\delta(\mathcal{L}^\alpha-\mathcal{L}^\alpha_{|\mathcal{R}\rangle})$ for $N\gg 1$, in agreement with Shnirelman's  theorem \cite{shnirel1974ergodic}.

In a genuine continuous Hamiltonian system with a smooth potential, obtaining a statistically meaningful number of consecutive eigenstates within a narrow energy window, for the study of distribution of localization measures, poses a significant challenge. Owing to the sparse matrix representation of quantum three-particle FPUT Hamiltonian, for the first time, we can employ the Krylov subspace method to obtain $N(E,\delta E, \hbar) \sim 1\times 10^3$ consecutive states of the singlet, with $\delta E \ll E$ for the statistics of the R\'enyi-Wehrl ELMs. And for the numerical calculation of ELMs, we need to discretize the Eqs. \eqref{eq:rw1}-\eqref{eq:elm}, so that
\begin{align}
    \label{eq:rw-d}
      H_k^{(\alpha)}&= \frac{1}{1-\alpha}\ln  \sum_{i,j} \left(\mathcal{Q}_k^{i,j}\right)^{\alpha},  \ \mathcal{L}_k^\alpha = \frac{\exp H_k^{(\alpha)}}{N_c},
\end{align}
where $\mathcal{Q}_k^{i,j}= \bar{\mathcal{Q}}_k^{i,j}/\sum_{i,j} \bar{\mathcal{Q}}_k^{i,j}$, and  $(i,j)$ indicates grid points $(q_{2,i},p_{2,j})\in\mathcal{S}_c$, $N_c$ is the number of grid points of the chaotic parts from SOS.

In Fig. \ref{fig:localizationHenon} we show the distributions of the $\alpha=1$ and $\alpha=2$ R\'enyi-Wehrl ELMs, for the $M\ge 0.8$ eigenstates, as a function of the Planck constant $\hbar$, at two energies for the singlet of $\alpha$-FPUT. The comparison is made with the best-fitting beta distribution in the main figures, complemented by an evaluation of cumulative distributions, as illustrated in the inset. At each energy, as $\hbar$ decreases, the agreement between $P(\mathcal{L}^\alpha)$ and the beta distribution improves. This improvement is characterized by reduced fluctuations and a decline in the proportion of ELMs with $\mathcal{L}^\alpha > \mathcal{L}^\alpha_{|\mathcal{R}\rangle}$, mostly ascribed to the influence of finite-size effects, because our numerical calculations involve the discretization of SOS with a grid size $\Delta q_2=\Delta p_2=5\times 10^{-3}$. A larger grid size (larger than the Planck cell) is equivalent to local averaging, resulting $\mathcal{L}^\alpha =1$ for the fully extended (chaotic) states. 
 
\begin{figure}
    \includegraphics[width=1\linewidth]{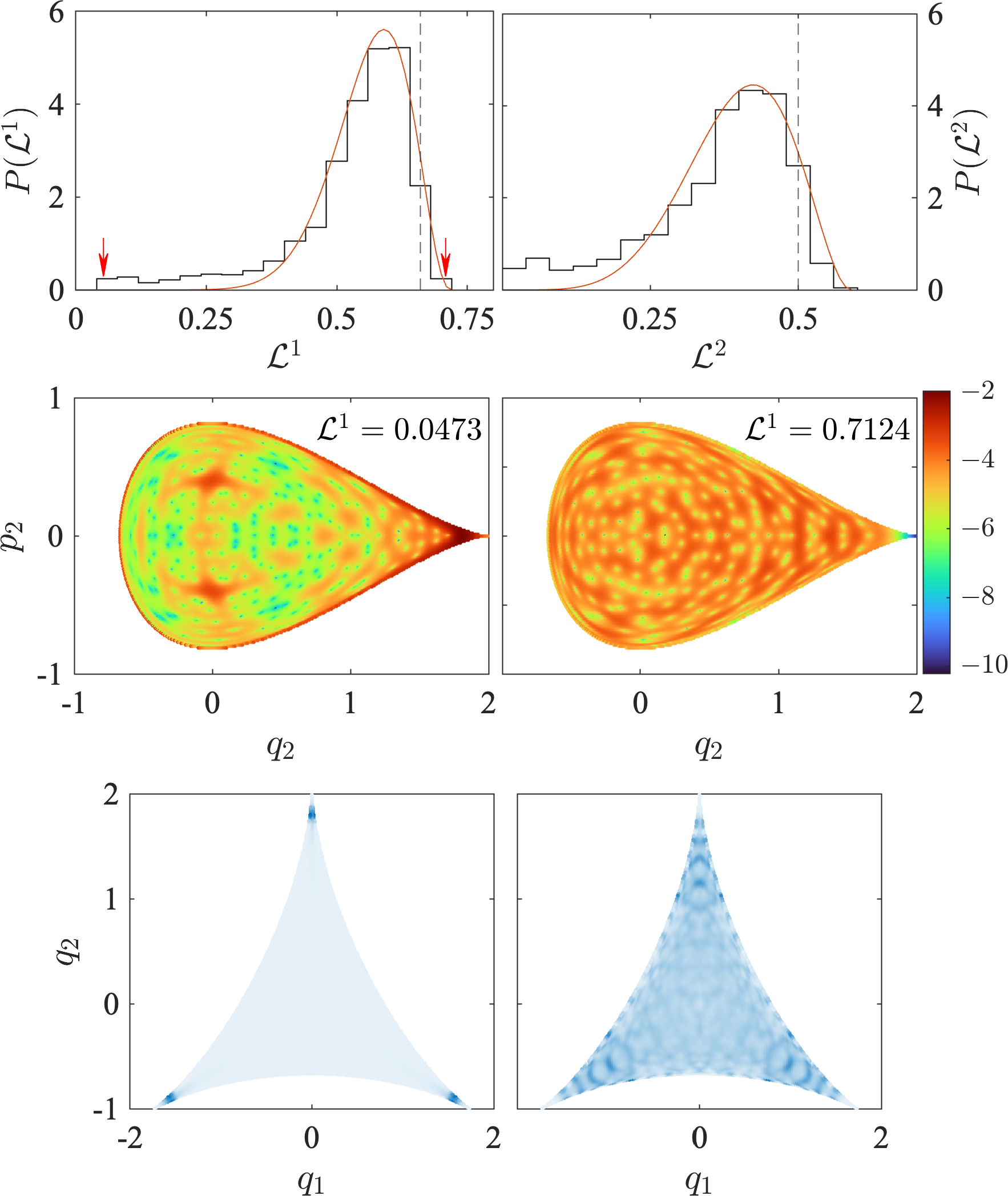}
    \caption{Top panels: distributions of the $\alpha=1$ and $\alpha=2$ R\'enyi-Wehrl ELMs of  $N(E,\delta E, \hbar) \simeq 1200$ eigenstates, from the singlet of general FPUT with $\lambda=1/16$, at energy $E=1/3$ with $\delta E/E\simeq 0.01$ and $\hbar=2\times 10^{-3}$. Parameters of the best-fitting beta distribution for $P(\mathcal{L}^1)$ are $(11.60, 2.55)$, $\mathcal{L}^1_{0}= 0.72$ and $\mathcal{L}^2_{0}= 0.59$. The vertical dashed lines indicate $\mathcal{L}^\alpha_{|\mathcal{R}\rangle}$. Middle panels: Husimi QSOSs given by Eq. \eqref{eq:QSOS-direct} (plotted in the logarithmic scale) of the maximally localized and extended states, which are pointed out by red arrows in the left top panel. Bottom panels: Projected Husimi functions in configuration space $\widetilde{\mathcal{P}}_k(q_1,q_2)$ given by Eq. \eqref{eq:husimiProjConfig} (plotted in the linear scale), for the same two states plotted previously in the middle panels.}
    \label{fig:localizationGen}
\end{figure}

At each energy, the value of the parameter $\beta_a$ in the best-fitting parameters for $P(\mathcal{L}^1)$ generally increases as $\hbar$ decreases, except in the case with $\hbar=8\times10^{-4}$ at $E=0.14$, where the fitting is less satisfactory. Moreover, the value of $\beta_a$ is higher at $E=0.16$ compared to $E=0.14$ for the same Planck constant, even when comparing the case of the former with $\hbar=9\times 10^{-4}$ to the latter with $\hbar=4\times 10^{-4}$. This observation can be explained by examining the controlling ratio $\alpha_\mathcal{L}$ as depicted in Fig. \ref{fig:alphaLocalization}($c$): In all cases, the ratio at $E=0.16$ is greater than that at $E=0.14$, resulting in better fittings and  a reduced degree of localization, characterized by the shrinking of the distributions. To provide additional evidence supporting the exclusive correlation between $\beta_a$ and $\alpha_\mathcal{L}$, we illustrate the distribution of ELMs in Fig. \ref{fig:localizationGen} (top panels), for the singlet of the general case with $\lambda=1/16$, at the saddle energy $E=1/3$. While the Planck constant $\hbar$ is set to $2\times10^{-3}$, a higher value compared to the $\alpha$-FPUT case where $\hbar$ was $4\times 10^{-4}$ at $E=0.16$, it's noteworthy that the parameter $\beta_a$ falls within the range observed for the latter with $\hbar=5\times 10^{-4}$ and that of $\hbar=4\times 10^{-4}$. This observation is explained by the findings depicted (denoted by colored dots) in Fig. \ref{fig:alphaLocalization}$(c)$, which illustrate that the value of $\alpha_\mathcal{L}$ is comparable in the two cases. In conclusion, the value of $\beta_a$ exhibits a monotonically increasing behavior with respect to $\alpha_{\mathcal{L}}$. As one expects, when $\alpha_{\mathcal{L}} \gg 1$, the trend suggests that $\beta_a$ tends towards infinity, but as unveiled in the study of quantum billiards \cite{batistic2019statistical}, this transition might be very slow. 

As discussed earlier, higher order R\'enyi-Wehrl ELMs amplify differences in probability densities, resulting in a broader distribution compared to the lower orders. Meanwhile, the more localized states in the lower tail of the distribution become more prominent, as illustrated in Figs. \ref{fig:localizationHenon}-\ref{fig:localizationGen}, where the difference between $P(\mathcal{L}^1)$ and $P(\mathcal{L}^2)$ of IPR is demonstrated. Having chosen the maximally localized and extended states from the distribution of $\alpha=1$ ELMs in the general case, as indicated in the top left panel of Fig. \ref{fig:localizationGen}, we then depict both the Husimi QSOS and projected Husimi functions for these two states. Remarkably, the results demonstrate that the maximally localized state concentrates around the saddle points, while the maximally extended state notably avoids these points, extending across the remaining space, complementing each other.  Identified as unstable (hyperbolic) fixed points, these saddles are elucidated in paper I, in polar coordinates $(r, \phi)$ as $(2, \phi_m)$, where $\sin 3\phi_m=-1$, $q_1=r\cos\phi$, $q_2=r\sin\phi$. Even more strikingly,  the Husimi QSOS of the maximally localized state shows that the localization is influenced by the stable and unstable manifold of these saddles, which exist in isolation and embedded in a sea of chaos, similar to the finding in Ref. \cite{waterland1988classical}. An extensive study of the classical periodic orbits and the stable and unstable manifolds of hyperbolic points, in relation to quantum scars, will be present in a forthcoming paper.

In the gallery of states presented in Fig. \ref{fig:HusimiGenGallery} in Appendix \ref{appD}, one can observe more examples of states, which are localized around the saddle points. It is this specific type of strong localization that contributes to the bulge, causing a deviation from the beta distribution in the lower tail. While most states notably avoid the saddles, displaying a much lesser degree of localization, some exhibit clear evidence of quantum scarring, indicating the presence of underlying unstable periodic orbits.

\section{Conclusions and Discussion}
\label{sec6}
In conclusion, this work presented a systematic study of the phenomenology of quantum eigenstates in the three-particle FPUT model, considering both the $\alpha$-type, which is canonically equivalent to the celebrated H\'enon-Heiles Hamiltonian, a nonintegrable and mixed type system, and the general case at the saddle energy where the system is fully chaotic (ergodic). While employing the circular (two-mode) basis for  quantization, we have derived a Wigner $d$-matrix decomposition of the unitary transformation from this circular basis to the Cartesian two-mode basis. The analytical expression of this unitary transformation plays a vital role for the discussion and comparison of various Husimi functions, with a particular focus on the completely projected Husimi function and the projected Husimi function from the energy shell in $(q_2, p_2)$ phase space. It is worth noting that the latter can be regarded as a blurred version of the former, as pointed out in Ref \cite{pilatowsky2021quantum}, and we make use of this effect to establish a connection (equivalence) between the projected Husimi function in configuration space and the probability density in configuration space  from the wavefunction.

Through a comprehensive examination of  Husimi functions, representing various eigenstates, we have identified the correspondence between   the Husimi QSOS and classical SOS. This correspondence establishes a valuable tool for distinguishing regular, mixed, and chaotic eigenstates in mixed-type systems. Furthermore, it allows for the quantification of  localization of the chaotic eigenstates. On the other hand, various projected Husimi functions prove to be effective tools for identifying highly localized states, and especially the periodic orbits underlying the quantum scarred states.

By introducing the overlap index $M$, a measure to quantify the overlap between the Husimi QSOS and the classical SOS, we define in mixed-type system, a quantity $\chi_M(E, \hbar)$,  the fraction of mixed eigenstates within an ensemble of states in an energy interval $[E-\delta E/2, E+\delta E/2]$, where $\delta E \ll E$. For states with $|M|\le 0.8$, i.e. states not entirely supported by either the regular islands or the chaotic region, there is a power-law decay concerning the decreasing value of $\hbar$, that $\chi_M(E,\hbar)\sim \hbar^\xi$. For states with $-0.8 \le M \le 0$, a power-law decay is also observed, yet with a slightly different power exponent. Moreover, the exponent varies with energy, while the relative phase space volume of the chaotic region varies with respect to energy, in the mixed-type system.

Upon examining various power-law decays,  it was observed that the mixed states with $M\le0$ exhibit a different decay rate than those with $M>0$. For mixed states with $M \leq 0$, the predominant mechanism involves (chaos-assisted) tunneling between regular islands, whereas for mixed states with $M>0$, the primary phenomenon is the flooding of chaotic states into regular regions. It delineates the disparities between these two tunneling mechanisms as one approaches the semiclassical limit. It is noteworthy to mention that within the gallery of states, we have identified also a recurring pattern of states with $M \simeq -1$. Integrating this observed pattern of $M \simeq -1$ states with the PUSC picture poses an intriguing open question for our future work.

We investigate the localization measure $\mathcal{L}^\alpha$ of chaotic eigenstates, as defined in terms of the R\'enyi-Wehrl entropy, in both mixed-type system and the fully chaotic system. $\mathcal{L}^\alpha$  is normalized exponential of the $\alpha$-R\'enyi-Wehrl entropy. While $\alpha=1$ refers to the Wehrl (information) entropy localization measure, the $\alpha=2$ case refers to the inverse participation ratio. The analysis is based on Husimi QSOS. For the first time, we explore the distribution of localization measures $P(\mathcal{L}^\alpha)$ for an ensemble of eigenstates from a narrow energy window,  in a genuinely continuous Hamiltonian system with a smooth potential. The degree of localization in the chaotic eigenstates is determined by the controlling ratio $\alpha_\mathcal{L} = t_H /t_T$, between the Heisenberg time $t_H$ and the classical transport time $t_T$. We have shown that a larger value of $\alpha_\mathcal{L}$ leads to a more accurate fit with a beta distribution of $P(\mathcal{L}^\alpha)$, where $\beta_a$ serves as the fitting parameter, and to its larger value. The monotonically increasing trend observed in the values of $\beta_a$ concerning $\alpha_{\mathcal{L}}$ implies that when $\alpha_{\mathcal{L}} \gg 1$, $\beta_a$ tends towards infinity, although this transition may be very slow. The condition $\beta_a\gg1$ leads to $P(\mathcal{L}^\alpha)=\delta(\mathcal{L}^\alpha-\mathcal{L}^\alpha_{|\mathcal{R}\rangle})$, where $\mathcal{L}^\alpha_{|\mathcal{R}\rangle}$ is the localization measure of the random pure state from RMT. This transition with respect to $\alpha_\mathcal{L} (\sim 1/\hbar)$ and the power-law decay of the mixed states with decreasing $\hbar$, together provide supporting evidence for the PUSC in the semiclassical limit. A theoretical analysis the of localization properties of chaotic eigenstates, especially a mathematical proof of the beta distribution describing the distribution of localization measures of chaotic eigenstates, is still an open question. We will leave this study for future work.

Moreover, we have found that the maximally localized state in the general case, which is a fully chaotic system, is influenced by the stable and unstable manifold of these saddles, which exist in isolation and embedded in a sea of chaos, while the maximally extended state notably avoids these points, extending across the remaining space, complementing each other. It is this specific type of strong localization that contributes to the bulge, causing a deviation from the beta distribution in the lower tail, which is expected to disappear in the strict semiclassical limit, according to PUSC. 

In the examination of projected Husimi functions, whether in configuration space or the phase space of a subspace, we have demonstrated the existence of distinct quantum scars.  An extensive study of the classical periodic orbits, including the stable and unstable manifolds of hyperbolic points, in relation to quantum scars, will be present in our forthcoming paper.

\section{Acknowledgement}
We express our gratitude to Dr. Jiaozi Wang for discussions and extend our appreciation to Matic Orel for his technical assistance. This work was supported by the Slovenian Research and Innovation Agency (ARIS) under the grant J1-4387.

  \appendix
  \section{Matrix elements in circular Fock basis for different coupling terms}
  \label{appA}
  Matrix elements in the circular basis of the cubic coupling terms are (for detailed derivation, see paper I)
  \begin{align}
      \label{eq:coeff-cubic}
    \langle n',l'|\hat{q}_\pm^3|n,  l\rangle=\hbar^{3/2}\delta_{l',l\pm3}\sum_{m\in\mathcal{M}_1} k_m^\pm (n,l)\delta_{n',n+m},
  \end{align}
  where $\mathcal{M}_1=\{\pm 1, \pm3\}$, $k_m^+(n,l)=k_m^-(n,-l)$, and 
  \begin{align}
  \begin{aligned}
      \label{eq:no-phase}
      &k_{-1}^+=3\sqrt{(n-l)(n-l-2)(n+l+2)/8}, \\
      &k_{1}^+=3\sqrt{(n-l)(n+l+2)(n+l+4)/8},  \\
      &k_{-3}^+=\sqrt{(n-l)(n-l-2)(n-l-4)/8},  \\
      &k_{3}^+=\sqrt{(n+l+2)(n+l+4)(n+l+6)/8}.
  \end{aligned}
\end{align}
For the quartic coupling, there is
  \begin{align}
    \langle n',l'|(\hat{q}_+\hat{q}_-)^2|n,l\rangle=\hbar^2\delta_{l'l}\sum_{m\in\mathcal{M}_2}k_m(n,l)\delta_{n',n+m}
  \end{align}
  where $\mathcal{M}_2=\{0, \pm 2, \pm4\}$, and the coefficients
  \begin{align}
  \begin{aligned}
      &k_0=\frac{3}{2}n^2-\frac{1}{2}l^2+3n+2, \\
      &k_{-2}= n\sqrt{n^2-l^2},\ k_2=(n+2)\sqrt{(n+2)^2-l^2},\\
      &k_{-4}=\frac{1}{4}\sqrt{n^2-l^2}\sqrt{(n-2)^2-l^2},\\
      &k_4=\frac{1}{4}\sqrt{(n+2)^2-l^2}\sqrt{(n+4)^2-l^2}.
  \end{aligned}
\end{align}
So in the circular basis $|n,l\rangle$, the cubic terms in quantum $\alpha$-FPUT couple states with $\Delta l=\pm 3$.  
Therefore, the cubic coupling takes place in three decoupled sets of basis states, two doublets and one singlet, agrees with the fact that $C_{3v}$ symmetry exists in the classical H\'enon-Helies potential. The quartic term in the  three-particle quantum $\beta$-FPUT does not introduce any coupling between different $l$, due to the conservation of angular momentum.
\begin{widetext}
\section{Unitary transformation between the Cartesian- and circular-mode}
  \label{appB}
  To get the unitary transformation between the circular-mode basis $|n,l\rangle $ and the Cartesian-mode basis $|n_1,n_2\rangle$,  first we expand Eq. \eqref{eq:fock-basis} by using the definition of rotated bosonic operator given in Eq. \eqref{eq:rotated-bosonic}
    \begin{align}
        \label{eq:unitary-m1}
       |n,l\rangle&= \sum_{n_1,n_2}\Omega_{n_1n_2}^{nl}|n_1,n_2\rangle
       =\frac{(a_+^{\dagger})^{n_+} (a_-^\dagger)^{n_-}}{\sqrt{n_+!n_-!}}|00\rangle= \frac{1}{2^{n/2}} \frac{(a_1^{\dagger}+ia_2^\dagger)^{n_+} (a_1^\dagger-ia_2^\dagger)^{n_-}}{\sqrt{n_+!n_-!}}|00\rangle\nonumber \\
    &=\sum_{k_1=0}^{\frac{n+l}{2}}\sum_{k_2=0}^{\frac{n-l}{2}}\frac{i^{k_1-k_2}}{2^{n/2}}\binom{\frac{n+l}{2}}{k_1}\binom{\frac{n-l}{2}}{k_2}\sqrt{\frac{(n-k_1-k_2)!(k_1+k_2)!}{(\frac{n+l}{2})!(\frac{n-l}{2})!}}
        |n-k_1-k_2,k_1+k_2\rangle.
    \end{align}
  Denoting $n=2j, l=2m$, one can rewrite Eq. \eqref{eq:unitary-m1} as 
  \begin{align}
     |n,l\rangle =|2j,2m\rangle
     =\sum_{k_1=0}^{j+m}\sum_{k_2=0}^{j-m}\frac{i^{k_1-k_2}}{2^{j}}\frac{\sqrt{(j+m)!(j-m)!(2j-k_1-k_2)!(k_1+k_2)!}}{k_1!k_2!(j+m-k_1)!(j-m-k_2)!}
      |2j-k_1-k_2,k_1+k_2\rangle.
  \end{align}
  Replacing $k_1+k_2=j-m'$, where $-j\le m'\le j$ obtained directly from $0 \le k_1+k_2
  \le 2j$, 
  \begin{align}
      |n,l&\rangle=\sum_{k_1=0}^{j+m}\sum_{k_2=0}^{j-m}\frac{i^{2k_1-j+m'}}{2^{j}}\frac{\sqrt{(j+m)!(j-m)!(j+m')!(j-m')!}}{(j+m-k_1)!k_1!(j-k_1-m')!(k_1-m+m')!}
      |j+m',j-m'\rangle,\nonumber \\
      &=\sum_{m'=-j}^j\sum_{k}\frac{i^{2k-j+m'}}{2^{j}}\frac{\sqrt{(j+m)!(j-m)!(j+m')!(j-m')!}}{(j+m-k)!k!(j-k-m')!(k-m+m')!}
      |j+m',j-m'\rangle.
  \end{align}
\end{widetext}
  The reordering of the double summation introduces a new range of $k$,  
  \begin{equation}
    \begin{aligned}
        k_{\min}&=\max(0,m-m'),\\
    k_{\max}&=\min(j-m',j+m). 
    \end{aligned}
  \end{equation}
  Thus, the unitary transformation can be further expressed as the following:
  \begin{align}
      |n,l\rangle  =\sum_{m'=-j}^{j}O_{m,m'}^j|j+m',j-m'\rangle,
  \end{align}
  where the coefficients $O_{m,m'}^j$ are expressed as
  \begin{align}
      O_{m,m'}^j=\sum_{k=k_{\min}}^{k=k_{\max}}& i^{2k-j+m'}[\cos(\pi/4)]^{2j-2k+m-m'}\nonumber\\
     &\times  [\sin(\pi/4)]^{2k-m+m'} W_k^{jmm'}
  \end{align}
  with 
  \begin{align}
     W_k^{jmm'}=\frac{\sqrt{(j+m)!(j-m)!(j+m')!(j-m')!}}{(j+m-k)!k!(j-k-m')!(k-m+m')!}.\nonumber
  \end{align} 

  Comparing with the Wigner's formula for Wigner $d$-matrix \cite{wigner2012group}
  \begin{align}
      d^j_{m,m'}(\theta)=\sum_k& (-1)^k \left(\cos\frac{\theta}{2}\right)^{2j-2k+m'-m}\nonumber\\ &\times \left(-\sin\frac{\theta}{2}\right)^{2k+m-m'}w_k^{jmm'},
  \end{align}
  where 
  \begin{align}
      w_k^{jmm'}=\frac{\sqrt{(j+m)!(j-m)!(j+m')!(j-m')!}}{(j-m-k)!k!(j-k+m')!(k+m-m')!},\nonumber
  \end{align}
  and $k\in[\max(0,m'-m),\min(j-m,j+m')]$, we get
  \begin{align}
      O_{m,m'}^j = i^{m'-j}d_{m',m}^j (-\pi/2)= i^{m'-j}d_{m,m'}^j(\pi/2),
  \end{align}
  using one of the symmetries $d_{m',m}^j (-\theta)=d_{m,m'}^j (\theta)$. So finally we have
  \begin{align}
      |n,l\rangle =\sum_{m'=-j}^j  i^{m'-j}d_{m,m'}^j(\pi/2)|j+m',j-m'\rangle.
  \end{align}
  
  Recently, it was found that the employing of Jacobi method can avoid the error and instability introduced by the summation over many terms with large numbers of alternating signs, based on the expansion of Wigner $d$-function in terms of the Jacobi polynomials \cite{Varshalovich1988Quantum},
  \begin{align}
      d_{m,m'}^j(\theta)=&\xi_{m,m'}\left[\frac{s!(s+\mu+\nu)!}{(s+\mu)!(s+\nu)!}\right]^{1/2}\nonumber \\
      &\times \left(\sin\frac{\theta}{2}\right)^\mu\left(\cos\frac{\theta}{2}\right)^\nu P_s^{(\mu,\nu)}(\cos\theta),
  \end{align}
  where $\mu,\nu$ and $s$ are related to $m,m'$ and $j$ by
  \begin{align}
      \mu=|m-m'|, \ \nu=|m+m'|,\ s= j-\frac{1}{2}(\mu+\nu),
  \end{align}
  and 
  \begin{align}
      \xi_{m,m'}=\begin{cases}
        1 &  \text{if \ } m'\ge m,\\
            (-1)^{m'-m} & \text{if \ } m'<m,
            \end{cases}
  \end{align}
where the Jacobi polynomial can be calculated by the recurrence relations. Here 
we generate all the elements of Wigner $d$-matrix straightly from another definition with
  \begin{align}
      d_{m,m'}^j (\theta)=\langle jm|e^{-i\theta J_y}|jm'\rangle,
  \end{align}
  where $J_y$ is one generator of the Lie algebra of SU(2) and SO(3), also a component of the angular momentum operator. The sparse matrix exponential (by employing the kernel polynomial method) would be more effective than the Jacobi method to generate the unitary matrix or the Wigner $d$-matrix for large $j$.

\section{Equivalent Husimi functions}
\label{appC}
The equivalence between two Husimi functions results directly from the following equation:
\begin{align}
    \label{eq:equiv}
    \langle \alpha_1,\alpha_2|n,l\rangle = \langle \alpha_+,\alpha_-|n,l\rangle,
\end{align}
given the coherent state $a_\pm|\alpha_\pm\rangle=\alpha_\pm|\alpha_\pm\rangle$, written as 
\begin{align}
   |\alpha_\pm\rangle = e^{-|\alpha_\pm|^2/2}e^{\alpha_\pm a_\pm^\dagger}|0_\pm\rangle,
\end{align}
where the vacuum two-mode state in two different representations $|0_+0_-\rangle\equiv |00\rangle$. Eq. \eqref{eq:equiv} can easily be verified from the following derivation: 
\begin{align}
    |&\alpha_+,\alpha_-\rangle = e^{-(|\alpha_+|^2+|\alpha_-|^2)/2}e^{\alpha_+ a_+^\dagger}e^{\alpha_- a_-^\dagger}|0_+0_-\rangle \nonumber\\
    &=e^{-(|\alpha_+|^2+|\alpha_-|^2)/2}e^{\alpha_+ a_+^\dagger+\alpha_- a_-^\dagger}|0_+0_-\rangle \nonumber\\
    &=e^{-(|\alpha_1|^2+|\alpha_2|^2)/2}e^{\alpha_1 a_1^\dagger+\alpha_2 a_2^\dagger}|00\rangle=|\alpha_1,\alpha_2\rangle,
\end{align}
Physically, it is a quite obvious result, because $|\alpha_+,\alpha_-\rangle$ and $|\alpha_1,\alpha_2\rangle$ correspond to the same point in phase space. 

The relationship between the projected Husimi function in configuration space $\widetilde{\mathcal{P}}_k(q_1,q_2)$, and the configuration-space probability density $|\psi_k(q_1,q_2)|^2$ from wavefunction can be elucidated by drawing connections to the Wigner function. The projected Husimi function is defined as in Eq. \eqref{eq:husimiProjConfig}
\begin{align}
        \widetilde{\mathcal{P}}_k(q_1,q_2)=\frac{1}{2\pi \hbar}\int \delta\left(E_k-H(\bf{q},\bf{p})\right)\mathcal{H}_k(\alpha_1,\alpha_2)dp_1dp_2. \nonumber
 \end{align}
Simultaneously,  the configuration-space probability density $|\psi_k(q_1,q_2)|^2$ can be expressed in terms of the Wigner function $W_k(\alpha_1,\alpha_2)$ as 
\begin{align}
    |\psi_k(q_1,q_2)|^2=\int_{-\infty}^{\infty} W_k(\alpha_1,\alpha_2) dp_1dp_2.
\end{align}
While the Husimi function $\mathcal{H}_k(\alpha_1,\alpha_2)$ can be understood as the Weierstrass transform of the Wigner function (smoothing by a Gaussian filter),
\begin{align}
    \mathcal{H}_k(\alpha_1,\alpha_2) = \frac{4}{\pi^2}\int &e^{-2|\alpha_1-\beta_1|^2}e^{-2|\alpha_2-\beta_2|^2}\nonumber\\
    &\times W_k(\beta_1,\beta_2)d^2\beta_1d^2\beta_2.
\end{align}

\begin{figure}
    \includegraphics[width=1\linewidth]{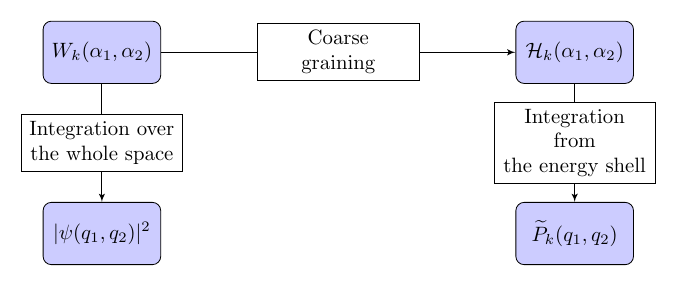}
    \caption{A flowchart illustrating the relation between the projected Husimi function in configuration space and the configuration-space probability density $|\psi_k(q_1,q_2)|^2$, linked through the Wigner function.}
    \label{fig:flowchart}
\end{figure}

\begin{figure*}
    \includegraphics[width=1\linewidth]{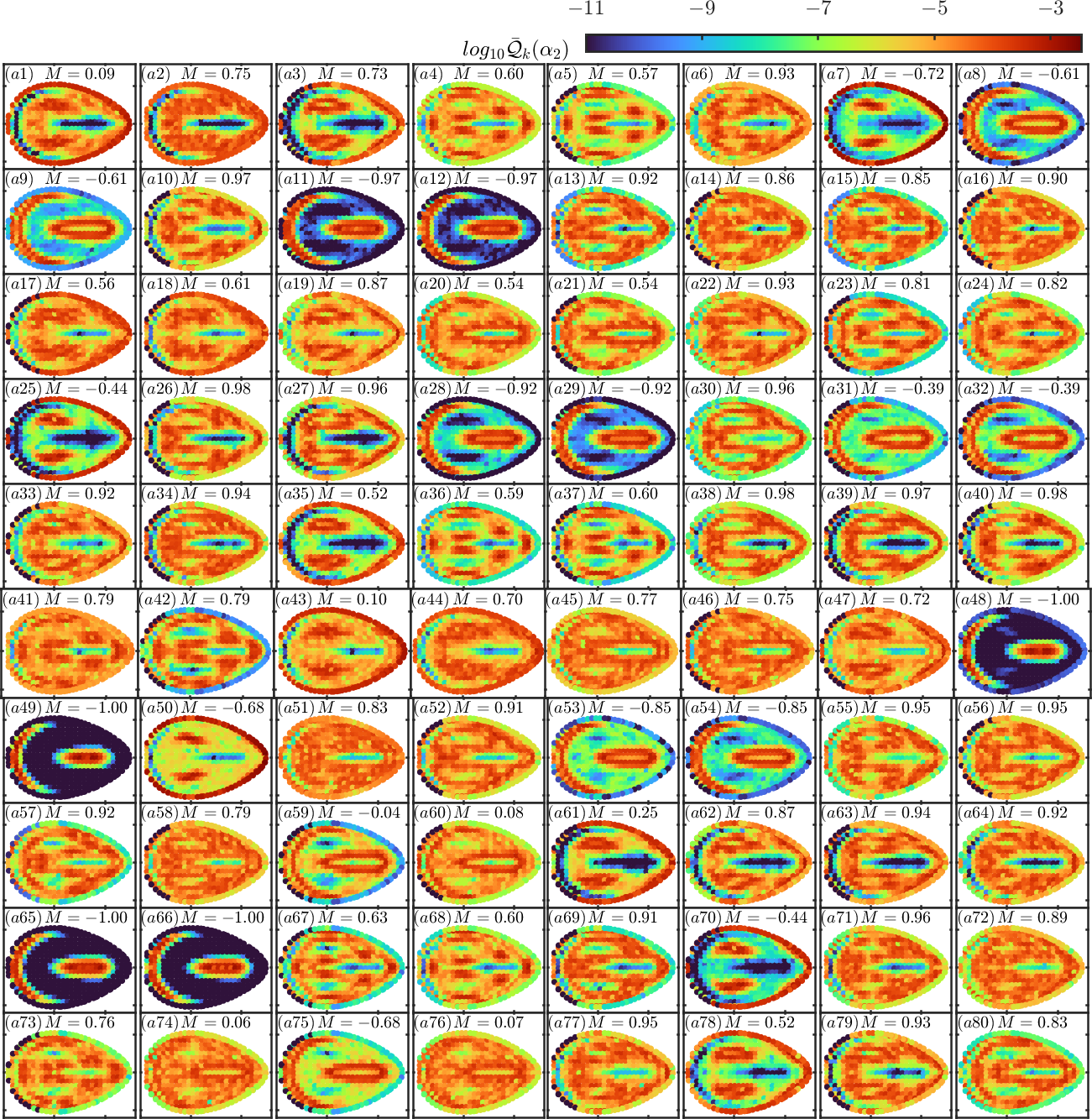}
    \caption{Examples of Husimi QSOS plotted in logarithmic scale, of 80 consecutive eigenstates in the energy interval $[E-\delta E/2,E+\delta E/2]$ at $E=0.14$ with $\delta E/E\simeq0.01$, for the singlet of $\alpha$-FPUT with $\alpha=1$, where the cutoff of $n$ is set to  $N=600$ and the Planck constant $\hbar=1\times 10^{-3}$.}
    \label{fig:henonGallery}
\end{figure*}

\begin{figure*}
    \includegraphics[width=1\linewidth]{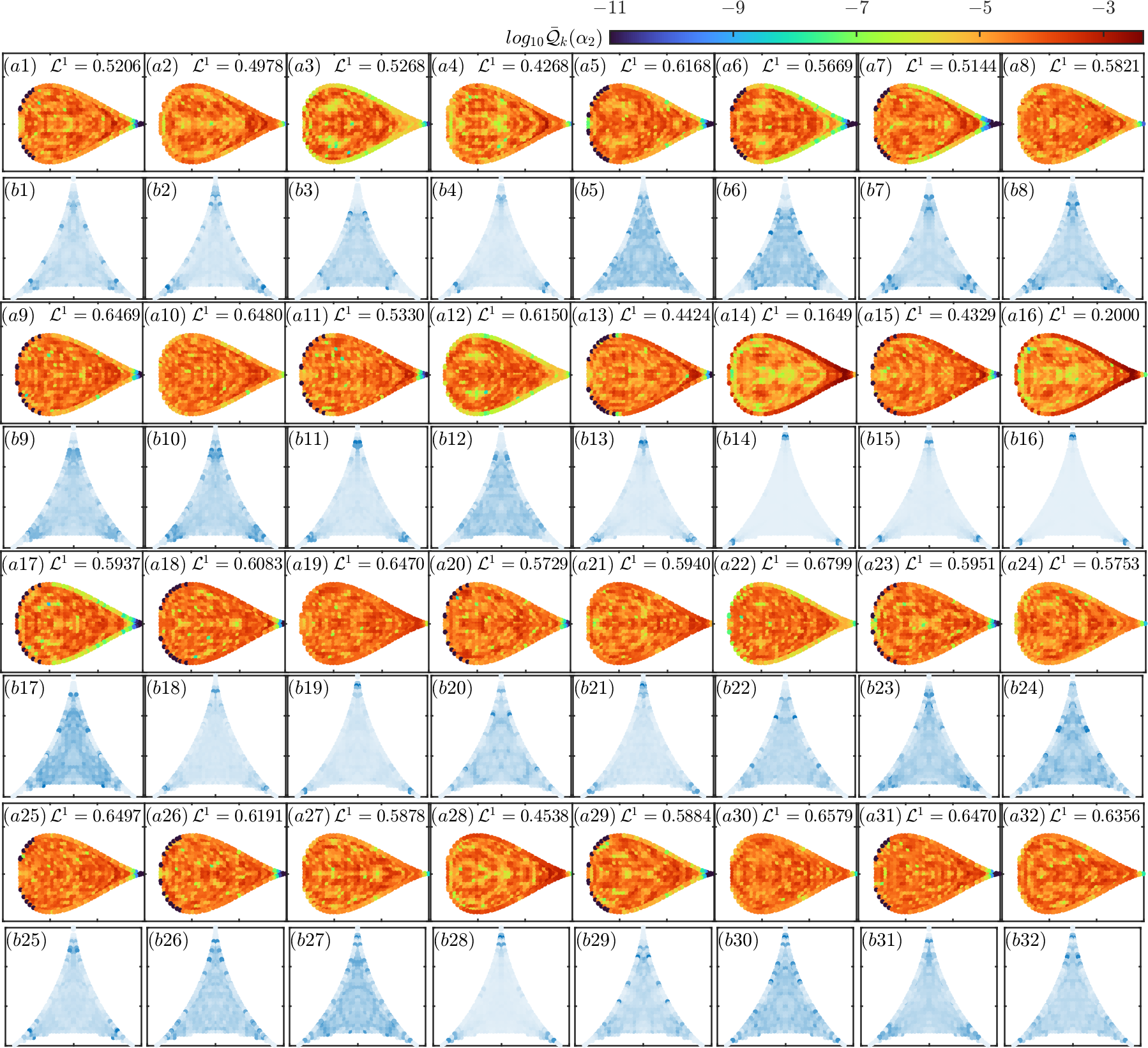}
    \caption{Examples of Husimi QSOS of Eq. \eqref{eq:QSOS-direct} plotted in logarithmic scale, and the corresponding projected Husimi functions in configuration space $\widetilde{\mathcal{P}}_k(q_1,q_2)$ given by Eq. \eqref{eq:husimiProjConfig} plotted in linear scale, of 32 consecutive eigenstates in the energy interval $[E-\delta E/2,E+\delta E/2]$ at $E=1/3$ with $\delta E/E\simeq0.0026$, for the singlet of  general FPUT with $\lambda=1/16$, where the cutoff of $n$ is set to $N=1200$ and the Planck constant $\hbar=2\times 10^{-3}$.}
    \label{fig:HusimiGenGallery}
\end{figure*}

Hence, $\widetilde{\mathcal{P}}_k(q_1,q_2)$ represents the integration over momentum within the energy shell $E_k$ of $\mathcal{H}_k(\alpha_1,\alpha_2)$. On the other hand, $|\psi_k(q_1,q_2)|^2$ corresponds to the integration over momentum across the entire momentum space of the Wigner function. The connection is visually depicted in the flowchart presented in Fig. \ref{fig:flowchart}. 
In our analysis of  $\widetilde{\mathcal{P}}_k(q_2,p_2)$ the  projection from energy shell and the completely projected Husimi function ${\mathcal{P}}_k(q_2,p_2)$  of Eq. \eqref{eq:completeProj}, as discussed in Section \ref{sec3.2}, we observed that integrating over the entire space leads to a blurring effect when compared to the integration specifically over the energy shell. This blurring effect can be interpreted as a form of coarse graining. Following the flowchart, one might reasonably conjecture that $\widetilde{\mathcal{P}}_k(q_1,q_2)$ is equivalent to $|\psi_k(q_1,q_2)|^2$, though lacking a rigorous mathematical proof.

\section{Gallery of states}
\label{appD}

In the main text, we have conducted a comprehensive phenomenological study of an ensemble of eigenstates, in an energy shell $[E-\delta E/2, E+\delta E/2]$ close to the energy surface, with $\delta E\ll E$. This investigation applies to both the mixed-type system and the fully chaotic system. One technical issue needs to be addressed here: the number of consecutive energy levels for the singlet is constrained, approximately determined by $N(E, \delta E, \hbar) \simeq g(E) \delta E/3$, where $g(E)$ is the semiclassical DOS. To ensure the convergence of the eigenspectra, the cutoff $N$, representing the limit of the quantum number $n$, needs to satisfy the condition that the total dimensions of the truncated Hilbert space, denoted as $\mathcal{N}=(N+1)(N+2)/2$, are significantly larger than $\mathcal{N}_E= \int g(E) dE$. In all our calculations in this work, $\mathcal{N}$ is ten or more times larger than $\mathcal{N}_E$, ensuring good convergence.

In Fig. \ref{fig:henonGallery}, we present examples of Husimi QSOSs given by Eq. \eqref{eq:QSOS-direct} for 80 consecutive eigenstates within the energy interval $[E-\delta E/2, E+\delta E/2]$ at $E=0.14$, where $\delta E/E \simeq 0.01$. These examples correspond to the singlet of the $\alpha$-FPUT system with $\alpha=1$, each is labeled with its respective overlap index $M$. It displays a distinct pattern illustrating the recurrence of states with $M \simeq -1$, as shown in Fig. \ref{fig:henonGallery}($a11$-$a12$), ($a28$-$a29$), ($a48$-$a49$), ($a65$-$a66$). The doubling can be explained that the singlet is a combination of $(A_1,A_2)$ symmetry, from the point symmetry group $C_{3\nu}$. While the phenomenon of repetitiveness has not been elucidated previously in quantum mixed-type systems, it may be associated with the same classical periodic orbit and can occur when the action difference satisfies $|\Delta I|=2\pi\hbar n$ $(n=1,2,\cdots)$, just like the scarring pattern in quantum chaotic systems \cite{gutzwiller2013chaos}. According to the PUSC, in the semiclassical limit, the states with $M\simeq-1$ would occupy a notable percentage within the ensemble of states if $1-\mu_c$ is not small. How to combine the PUSC picture with the $M\simeq-1$ pattern, will be an exciting open problem for our future works. In Fig. \ref{fig:HusimiGenGallery}, we show the gallery of states of both Husimi QSOSs of Eq. \eqref{eq:QSOS-direct}, each is labeled with its respective $\alpha=1$ ELM and projected Husimi functions  in configuration space given by Eq. \eqref{eq:husimiProjConfig}, for the general FPUT with $\lambda=1/16$, at saddle energy $E=1/3$. There are basically two types of states: those that concentrate around  the saddle points, exhibiting strong localization, and those that avoid the saddles, displaying a lesser degree of localization, showing clear evidence of quantum scarring, indicating underlying unstable periodic orbits.
\bibliographystyle{apsrev4-2}
\bibliography{yr2.bib}

\end{document}